\documentclass[a4paper,11pt]{article}
\pdfoutput=1 

\usepackage{jheppub} 

\usepackage[T1]{fontenc} 
\pdfoutput=1
\usepackage{graphicx,color,rotating}
\usepackage{hyperref}
\usepackage{epsfig,color}
\usepackage{slashed}
\usepackage{tabularx}
\usepackage{graphicx}
\usepackage{adjustbox}
\usepackage{tabularx}

\newcommand{\fr}[2]{{\hbox{$ #1 \over #2 $}}}

\bigskip

\title{Annihilation signatures of neutron dark decay models in neutron oscillation and proton decay searches}

\author{
Wai-Yee Keung,$^{1,2}$  Danny Marfatia,$^3$ and
Po-Yan Tseng$^{4,2}$}
\affiliation{
$^1$ Department of Physics, University of Illinois at Chicago,
Illinois 60607 USA \\
$^2$ Physics Division, National Center for Theoretical Sciences,
Hsinchu, Taiwan \\
$^3$ Department of Physics and Astronomy, University of Hawaii at Manoa,
Honolulu, HI 96822, USA \\
$^4$Kavli IPMU (WPI), UTIAS, The University of Tokyo, 
Kashiwa, Chiba 277-8583, Japan
}
\date{\today}

\abstract{
We point out that two models that reconcile the neutron lifetime anomaly via
dark decays of the neutron, also predict dark matter-neutron ($\bar{\chi}-n$) annihilation
that may be observable in  neutron-antineutron oscillation and proton decay searches
at Super-Kamiokande, Hyper-Kamiokande and DUNE.
We study signatures of $\bar{\chi}n\to \gamma\pi^0$~(or multi-$\pi^0$) and
$\bar{\chi}n\to \phi\gamma\pi^0$~(or $\phi+$multi-$\pi^0$), where $\phi$ is an almost massless boson in one of the two models. 
}

\begin{document}
\maketitle
\flushbottom

\section{Introduction}

In the standard model (SM), the neutron almost exclusively decays through beta decay, 
$n \to p+e^-+\bar{\nu}_e$, with Br$(n \to p +{\rm anything})=1$.
The neutron lifetime is measured in bottle experiments and beam experiments which
use different methodologies. In bottle experiments, the total neutron lifetime $\tau_n^{\rm bottle}$ 
is  measured 
by counting the number of neutrons trapped in a container as function of time.  On the other hand, 
beam experiments count the number of protons resulting from neutron decay in a neutron beam.
In this case, the neutron lifetime is given by
$$
\frac{1}{\tau_n^{\rm beam}}=-\frac{{\rm Br}(n \to p +{\rm anything})}{N_n}\frac{dN_n}{dt}\,,
$$ 
where $N_n$ is the number of neutrons in the beam.
In the SM, the two methods should give the same neutron lifetime.
However, there is tension between the bottle~\cite{Pichlmaier:2010zz,Steyerl:2012zz,Arzumanov:2015tea} and 
beam~\cite{Byrne:1996zz,Yue:2013qrc} measurements of the neutron lifetime
at about the $4\sigma$ level~\cite{Patrignani:2016xqp}:
\begin{eqnarray}
\tau_n^{\rm bottle}&=& 879.6\pm 0.6 \ {\rm s}\,, \nonumber \\
\tau_n^{\rm beam}&=& 888.0\pm 2.0 \ {\rm s}\,. \nonumber
\end{eqnarray}
If neutrons decay through channels without protons in the final state i.e., Br$(n \to p +{\rm anything})<1$,
beam experiments will measure a longer lifetime than bottle experiments since
$$
\tau_n^{\rm beam}=\frac{\tau_n^{\rm bottle}}{{\rm Br}(n \to p +{\rm anything})}\,.
$$

To explain the discrepancy, the neutron decay width into channels without protons must be
$$
\Delta \Gamma(n \to {\text {no proton}})\simeq 7.1\times 10^{-30}~{\rm GeV}\,,
$$
which is about 1\% of the total neutron decay width.

Two models, dubbed Model~I and Model~II, invoking dark decays of the neutron were proposed in 
Refs.~\cite{Fornal:2018eol,Fornal:2018ubn}.
The basic idea is to introduce a tiny mixing between the neutron and a new particle 
in the dark sector,  
which allows the neutron to decay into dark sector particles  $\chi$ (a Dirac fermion) and $\phi$
(a scalar)  through
$n \to \chi+\gamma$ and $n \to \chi+\phi$.
The $\chi$ couples very weakly with the SM sector so as to not
trigger a detectable signal in the beam experiments.
Meanwhile, the mass of $\chi$ is restricted in a narrow window,
$$
937.992~{\rm MeV}< m_\chi < 938.783~{\rm MeV}\,,
$$
to simultaneously satisfy the requirement of $^9$Be stability 
and prevent the decay, $\chi \to p+e^-+\bar{\nu}_e$~\cite{Fornal:2018eol,Fornal:2018ubn,Pfutzner:2018ieu}.
These criteria make $\chi$ a good dark matter (DM) candidate
if other decay modes in the dark sector are forbidden. 
Based on these assumptions,
$\bar{\chi}-n$ annihilation provides detectable signals 
with energy of $\mathcal{O}(\rm GeV)$ 
in Super-Kamiokande (Super-K) 
or in the future experiments, Hyper-Kamiokande (Hyper-K) and DUNE.
The signals are similar to that for proton decay and neutron-antineutron oscillations. 
For instance in Model I, prompt photon and multi-$\pi^0$ signals are obtained 
from $\bar{\chi}n$ annihilation.
Dark matter-nucleon annihilation has been studied in a broader context in Ref.~\cite{Jin:2018moh}.

Note that these models have trouble with neutron star stability~\cite{McKeen,Baym,Motta}.
The conversion of neutrons to dark matter in the neutron star softens the nuclear equation of state to the
point that neutron stars above two solar masses are not possible, which is in contradiction with observations.
In a recent paper~\cite{grinstein} it was shown that extending Model~II with strong repulsive
DM-baryon interactions solves the problem. This extension does not contribute additional diagrams to the annihilation signatures we discuss. We view the models of Refs.~\cite{Fornal:2018eol,Fornal:2018ubn,grinstein} as examples that produce neutron dark decays
that may be extended with complex dark sectors to address various experimental and astrophysical constraints.


The paper is organized as follows. 
In section\,\ref{sec:int}, we present the relevant effective interactions 
and parameterize the required form factor. We introduce Model~I and -II 
for the dark decays of the neutron in section\,\ref{sec:m1} and \ref{sec:m2}, respectively.
Signatures of DM-neutron annihilation are discussed in section\,\ref{sec:cs},
and the signal event numbers at underground experiments are estimated in section\,\ref{sec:signal}.
We summarize our results in section\,\ref{sec:result}. We describe some of our cross section calculations in an Appendix.

\section{Effective interactions and form factor}
\label{sec:int}

Since the neutron has no electric charge, it couples to the photon via
the magnetic dipole interaction,
\begin{equation}
\label{eq:nan}
\mathcal{L}^{\rm eff}\supset \frac{g_n e}{2m_n}\, F_{\bar{n}\gamma n}(Q^2)\, \bar{n}\sigma^{\mu\nu}F_{\mu\nu}n\,,
\end{equation}
where $g_n\simeq -3.826$ is the neutron $g$-factor  
and $F_{\bar{n}\gamma n}(Q^2)$ is the corresponding form factor.
On the other hand, the pion-neutron and pion-proton effective interactions
satisfy the isospin symmetry and  are given by
\begin{eqnarray}
\label{eq:npin}
\mathcal{L}^{\rm eff}&\supset &
\frac{g_{n\pi}}{\sqrt{4\pi}}\,F_{\bar{n}\pi n}(Q^2)
\,\bar{N}(\overrightarrow{\tau}\cdot \overrightarrow{\pi})i\gamma_5 N                 \nonumber \\
&=&
\frac{g_{n\pi}}{\sqrt{4\pi}}\,F_{\bar{n}\pi n}(Q^2)
\,\left(
-\bar{n}i\gamma_5 n \pi^0 + \bar{p}i\gamma_5 p \pi^0 
+ \sqrt{2}\bar{p}i\gamma_5 n \pi^+ + \sqrt{2}\bar{n}i \gamma_5 p \pi^-
 \right)\,, \nonumber
\end{eqnarray}
where $N=(p,n)^T$, 
$\overrightarrow{\tau}\cdot \overrightarrow{\pi}=\sqrt{2}(\tau_-\pi_-+\tau_+\pi_+)+\tau_3 \pi_0$, $g_{n\pi}=\sqrt{13.54}$~\cite{deSwart:1997ep}, 
and $F_{\bar{n}\pi n}(Q^2)$ is the form factor.
%
For the $\bar{n}\pi n$ vertex with only $\bar{n}$ off-shell at 
momentum squared $-Q^2$, we parameterize the form factor as
\begin{equation}
\label{eq:ff}
F_{\bar{n}\pi n}(Q^2)
=\left(\frac{1-m_n^2/\Lambda_n^2}{1+Q^2/\Lambda_n^2}\right)^y\,,
\end{equation}
where $y$ is an unknown exponent and will be treated as a free parameter determined by data.
$y=0$ represents the case of no form factor suppression,
while $y=2$ gives a good fit to 
the electromagnetic form factor of the proton.
We determine the value of $y$  by comparing with the experimentally measured
$\bar{n}p$ annihilation cross section 
in the nonrelativistic limit. 
Here, $\Lambda_n$ characterizes the size of the nucleon, and is typically
$4\pi f_\pi \approx 1.2$~GeV.
The form factor is normalized to unity for $-Q^2=m_n^2$.
Furthermore, we  assume the same behavior for the magnetic form factor  
of the neutron. 
The complete amplitude also involves the transitional vertices between 
$\bar\chi-n$. 
%
%
We suppose that the form factors
$F_{\bar{\chi}\gamma  n}(Q^2)$, $F_{\bar{\chi}\pi n}(Q^2)$ and
$F_{\bar{n}\pi n}(Q^2)$, all have the same parameterization but may have different values for the exponent $y$, and $y=2$ for
the electromagnetic form factor.

\section{Model~I}
\label{sec:m1}

Model~I allows the dark decay $n \to \chi+\gamma$ by
introducing  two dark sector particles, 
a Dirac fermion $\chi$ 
(whose antiparticle $\bar{\chi}$ we identify as the DM candidate)
and a heavy scalar mediator $\Phi$ 
(color triplet, weak singlet, hypercharge $\fr{Y}2=-\fr13$).
The new interaction Lagrangian terms that contribute to $\chi-n$ mixing are~\cite{Fornal:2018eol}
$$
\mathcal{L}_1\supset \lambda_1 \Phi^*\chi d_R+\lambda'_1 \Phi u_R d_R+ {\text{h.c}}.
$$
As stated earlier,
$$
937.992~{\rm MeV}< m_\chi < 938.783~{\rm MeV}\,.
$$
The colored $\Phi$ must be much heavier than 1~TeV to be compatible with LHC data. 
It is therefore reasonable to work in the effective theory framework to describe 
processes at the GeV scale.
The DM-triquark operator is derived by integrating out the heavy scalar $\Phi$:
$$
\mathcal{L}\subset \frac{\lambda_1\lambda'_1}{m^2_\Phi}(\chi u_R d_R d_R)
=\frac{\lambda_1\lambda'_1}{m^2_\Phi}\beta (\chi n)\,,
$$
where form factor $\beta = \langle 0|u_Rd_Rd_R|n \rangle \simeq 0.0144~{\rm GeV}^3$~\cite{Aoki:2017puj}. 
This operator is effectively an off-diagonal mass term between the DM and neutron, leading to a $m^2_\Phi$ suppressed mixing angle,
\begin{equation}
\label{eq:angle}
\theta \simeq \frac{\varepsilon}{m_n-m_\chi}\,,
\end{equation} 
where $\varepsilon\equiv \beta \lambda_1\lambda'_1/m^2_\Phi$.

\begin{figure}[t!]
\centering
\includegraphics[height=2.0in,angle=0]{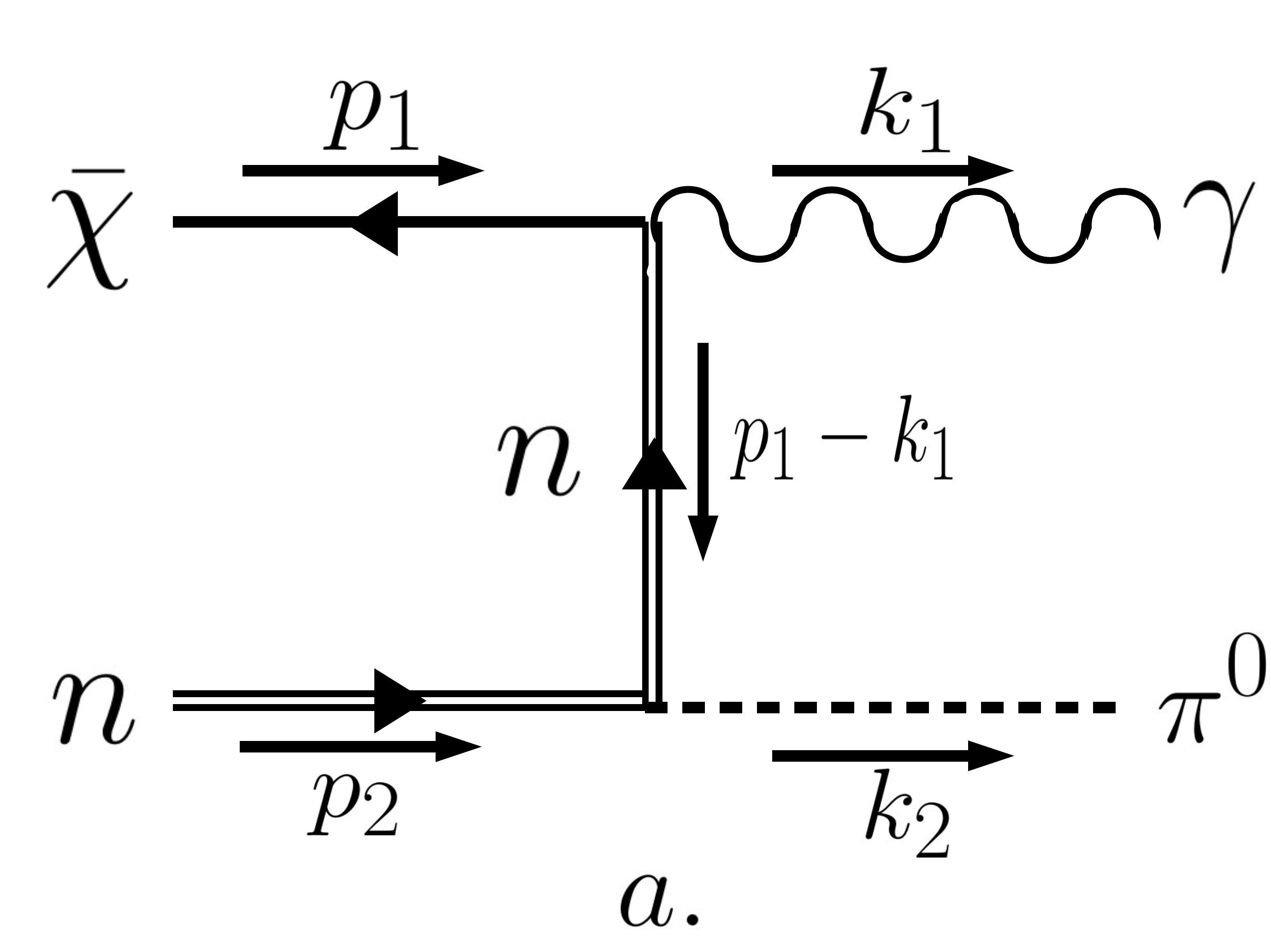} \hbox to 0.5in{}
\includegraphics[height=2.0in,angle=0]{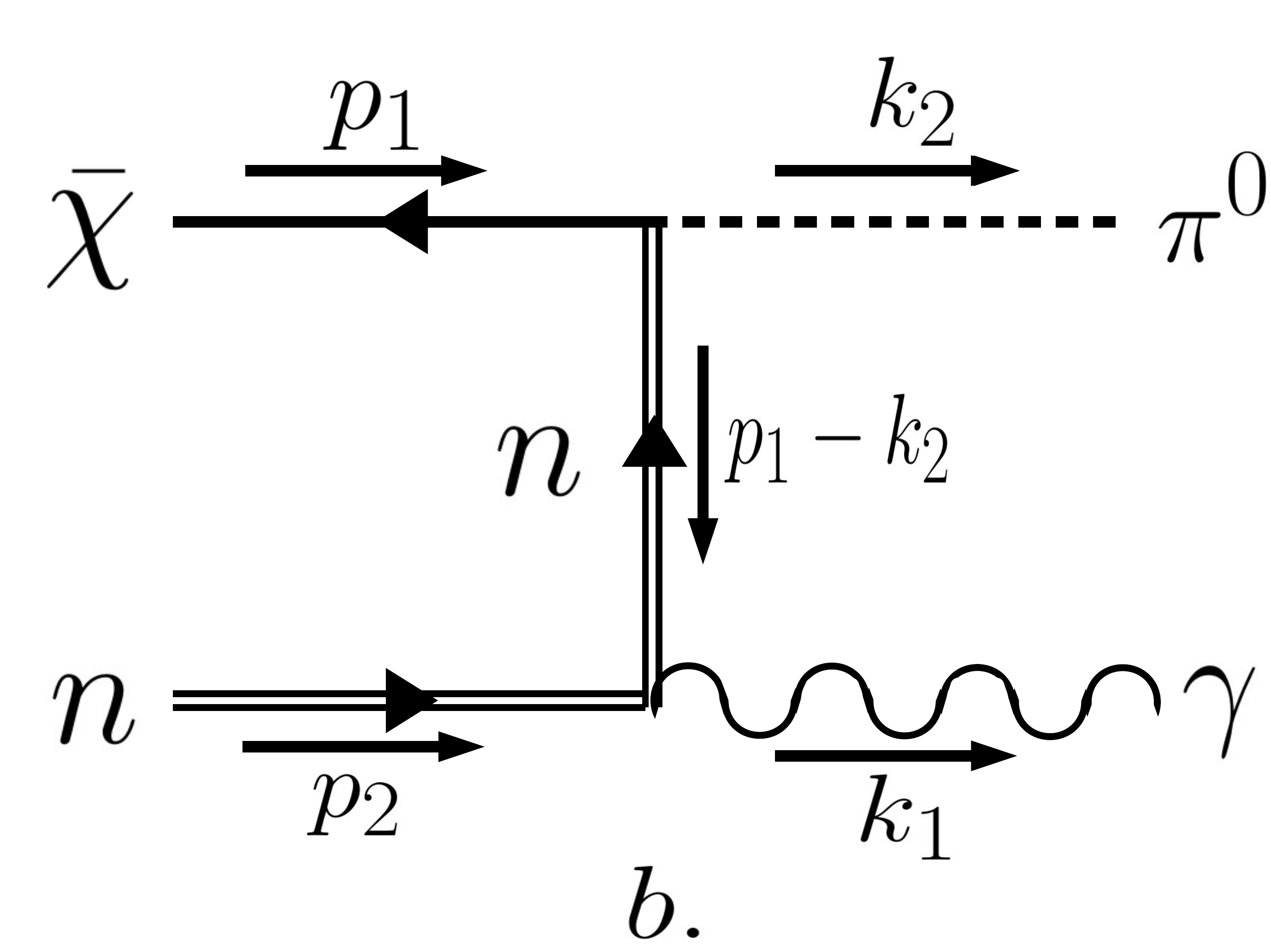}
\caption{\small \label{f1-1} 
Feynman diagrams for $\bar{\chi}+n \to \gamma+\pi^0$ in Model~I.
}
\end{figure}

The DM-neutron mixing gives rise to a DM-neutron-photon coupling,
which is responsible not only for neutron dark decay $n\to \chi \gamma$,
but also predicts the DM-neutron annihilation channel,
$$
\bar{\chi}+n \to \gamma+\pi^0\,, \nonumber
$$
as shown in Fig.~\ref{f1-1}.
A nonrelativistic DM particle in the halo interacts with 
a static neutron target and produces a photon and pion back-to-back, each with an energy of about a GeV. 
The topology of the signal is similar 
to that of the proton decay channel $p\to e^+\pi^0$ at experiments like Super-K. 
Both produce 3 electron-like Cherenkov rings in Super-K~\cite{Miura:2016krn}, and the reconstructed total 
momentum $P_{\rm tot}$ tends to be small.
However, the reconstructed invariant mass $M_{\rm tot}\simeq 2$\,GeV from $\bar{\chi}+n \to \gamma+\pi^0$,
is higher than $M_{\rm tot}\simeq 1$\,GeV from $p\to e^+\pi^0$.

A multi-pion final state can also result from DM-neutron annihilation.
It is analogous to $\bar{n}$-nucleon annihilation in the SM which predominantly
yields a multi-pion final state.
The DM-nucleon annihilation cross section is related to the 
$\bar{n}$-nucleon cross section via~\cite{Jin:2018moh}
$$
\sigma(\bar{\chi}N \to {\text{multi-$\pi^0$}})=\theta^2\, \sigma(\bar{n}N \to {\text{ multi-$\pi^0$}})\,. \nonumber
$$
This multi-$\pi^0$ signal can be detected in Super-K as well,
and the signal kinematics is similar to that for
$n-\bar{n}$ oscillations~\cite{Abe:2011ky}.
However, because of the compositeness of the hadron 
and a lack of experimental measurements of the $\bar{n}\pi n$ form factor, 
a perturbative calculation using Eqs.~(\ref{eq:nan}),~(\ref{eq:npin}) 
cannot give a precise estimate of the annihilation cross section.
%
%

To overcome this, we match our calculation to the experimentally measured $\bar{n}p$ annihilation cross section~\cite{Mutchler:1988av,Feliciello:1999ti,Bertin:1997gn,Armstrong:1987nu}
and multiply it by the mixing angle square $\theta^2$ to obtain the $\bar{\chi}-n$ annihilation cross section. We elaborate on our procedure in section~\ref{sec:cs}.

\section{Model~II}
\label{sec:m2}

Model~II has a richer structure than Model~I with two additional dark sector particles: 
a Dirac fermion $\tilde{\chi}$ and a complex scalar $\phi$~\cite{Fornal:2018eol}.
After the heavy scalar $\Phi$ is integrated out, 
$\tilde\chi$ mixes with the neutron through the mixing angle
\begin{equation}
\label{eq:angle2}
\theta\simeq \frac{\varepsilon}{m_n-m_{\tilde{\chi}}}\,,
\end{equation}
which is the same as Eq.~(\ref{eq:angle}) with $m_\chi$ replaced by $m_{\tilde{\chi}}$.
Then $\chi$ couples to $\phi$ and $\tilde{\chi}$
via the new interaction,
$$
\label{eq:l2}
\mathcal{L}\subset \lambda_\phi \bar{\tilde{\chi}}\chi \phi+{\rm h.c.} \nonumber
$$  
So, in addition to $n\to \gamma \tilde{\chi}$, 
a new neutron dark decay channel $n \to \phi\chi$ is allowed.
Hence, the sum of the decay widths,
$
\Delta \Gamma_{n\to \gamma\tilde{\chi}}+\Delta \Gamma_{n\to \phi\chi}
\simeq 7.1\times 10^{-30}~{\rm GeV}\,,
$
to reconcile the tension between beam and bottle experiments.

For $m_\chi > m_\phi$, 
the the annihilation channel $\bar{\chi}\chi \to \phi\phi$ 
via $t$-channel $\tilde{\chi}$ exchange can provide the correct DM relic density
if $\lambda_\phi\simeq 0.04$.
The three masses $m_\chi$, $m_\phi$, and $m_{\tilde{\chi}}$ 
should satisfy the relations,
\begin{eqnarray}
&& 937.992~{\rm MeV} < m_\chi+m_\phi < 939.565~{\rm MeV}\,, \nonumber \\
&& 937.992~{\rm MeV} < m_{\tilde{\chi}}\,, \nonumber \\
&& |m_\chi-m_\phi|< m_p+m_e = 938.783081~{\rm MeV}\,, \nonumber
\end{eqnarray}
to prevent $^9$Be decays to 
$^8{\rm Be}+\chi+\phi$ and
$^8{\rm Be}+\tilde{\chi}$, 
and to prohibit $\chi \to p+e^-+\bar{\nu}_e$, respectively~\cite{Fornal:2018ubn,Pfutzner:2018ieu}.
We choose three benchmark points with $\lambda_\phi\simeq 0.04$:
\begin{eqnarray}
{\bf P1}: ~~&& (m_\chi,m_\phi,m_{\tilde{\chi}})
=(937.992,0,937.992)  \nonumber \\
{\bf P2}: ~~ && (m_\chi,m_\phi,m_{\tilde{\chi}})
=(937.992,0,2m_n)  \nonumber \\
{\bf P3}: ~~ && (m_\chi,m_\phi,m_{\tilde{\chi}})
=(939.174,0.391,940.000)\,,  \nonumber 
\end{eqnarray}
where $m_\chi\, m_\phi$ and $m_{\tilde{\chi}}$ are in MeV.
All three points explain the neutron lifetime anomaly 
with the corresponding values of $\theta$ listed in Table\,\ref{tab-1}.
{\bf P1} and {\bf P2}, respectively, are the points from
Refs.~\cite{Fornal:2018eol,Fornal:2018ubn}, 
with $n\to \tilde{\chi}\gamma$ kinematically allowed 
for {\bf P1} but not for {\bf P2}. 
Since $\tilde{\chi}$ plays the role of a propagator in the DM-neutron annihilation process,
the signal event distributions are different for {\bf P1} and {\bf P2}.
For {\bf P3}, the DM-neutron annihilation cross section is maximized,
as we will see in the next section.

\begin{figure}[t!]
\centering
\includegraphics[height=2.0in,angle=0]{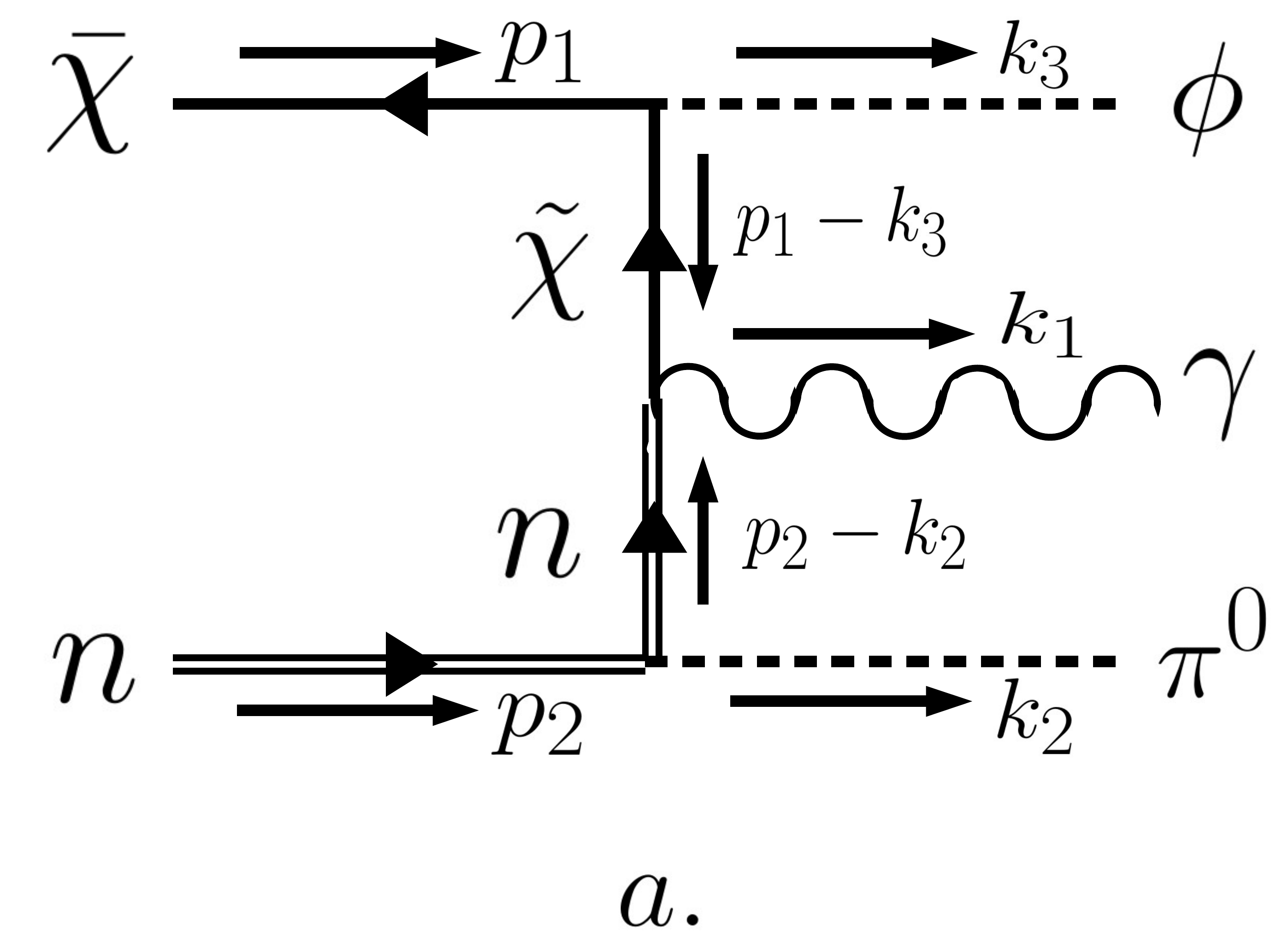} \hbox to 0.5in{}
\includegraphics[height=2.0in,angle=0]{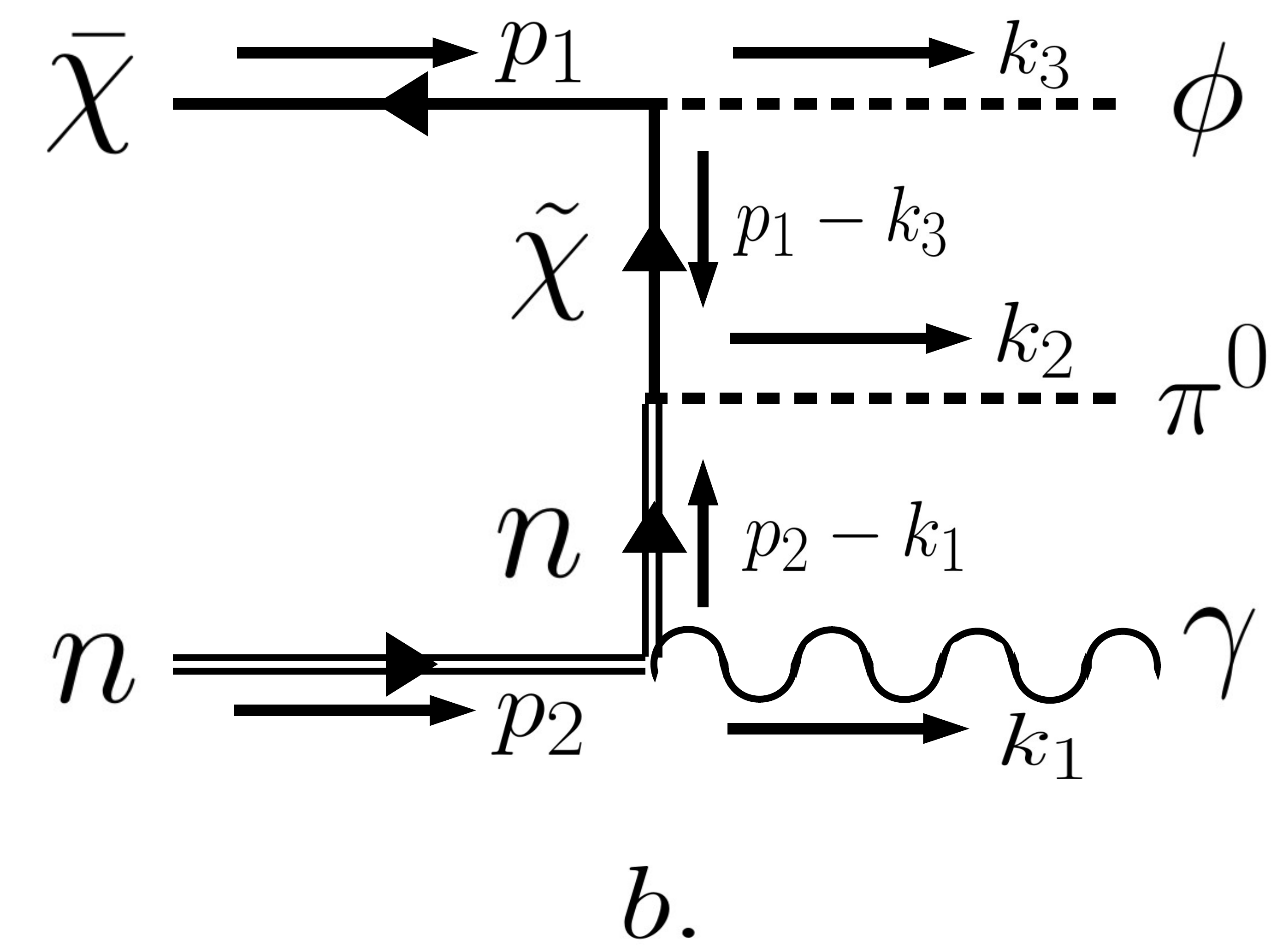}
\caption{\small \label{f1-3} 
Feynman diagrams for $\bar{\chi}+n \to \phi+\gamma+\pi^0$ in Model~II.
}
\end{figure}

Feynman diagrams for the DM-neutron annihilation process,
$$
\bar{\chi}+n \to \phi+\gamma+\pi^0\,, \nonumber
$$
are shown in Fig.~\ref{f1-3}.
The event distributions will be different from Model~I, 
due to the additional dark sector particle $\phi$ in the final state.
Since $\phi$ can escape the detector, 
the reconstructed invariant mass $M_{\rm tot}$ 
and total momentum $P_{\rm tot}$ from $\gamma$ and $\pi^0$
have different distributions from Model~I.

\section{Dark matter-nucleon annihilation cross section}
\label{sec:cs}

  \begin{table}[thb!]
\caption{\small \label{tab-1}
Input parameters and observables for two benchmark points for Model I and points {\bf P1}, {\bf P2} and {\bf P3} for Model II. Experimental cuts have not be applied.
}
\begin{adjustbox}{width=\textwidth}
\begin{tabular}{c|cc|ccc}
\hline
\hline
 & \multicolumn{2}{c |}{\bf Model I} & {\bf P1} & {\bf P2} & {\bf P3}  \\
\hline
\hline
$m_\chi$ [MeV]    & 937.992 & 938.783  & 937.992 & 937.992 & 939.174  \\
$m_\phi$ [MeV]    & -   & - & 0 & 0 &  0.391  \\
$m_{\tilde{\chi}}$ [MeV]    & -   & - & 937.992 & 2$m_n$ & 940.000  \\
$\lambda_\phi$    & -   & - & 0.04  & 0.04 & 0.04  \\
\hline
$|\theta|$
  &  $6.14\times 10^{-11}$ &  $1.75\times 10^{-10}$  
  & $4.21\times 10^{-12}$  & $4.21\times 10^{-12}$ & $4.03\times 10^{-11}$  \\
  $\Gamma_{n \to \chi \gamma~({\rm or}~\tilde{\chi} \gamma)}$ [GeV]    
  & $7.1\times 10^{-30}$   & $7.1\times 10^{-30}$ 
  & $3.3\times 10^{-32}$  & 0 & 0  \\
  $\Gamma_{n \to \chi \phi}$ [GeV]    
  & -   & - 
  & $7.07\times 10^{-30}$  & $7.10\times 10^{-30}$ & $7.10\times 10^{-30}$  \\
\hline
& \multicolumn{2}{c |}{$\bar{\chi}n \to \gamma \pi^0~(y=2)$} & \multicolumn{3}{c}{$\bar{\chi}n\to\phi\gamma \pi^0~(y=2)$} \\
\hline
$\frac{v}{c}\sigma$ [cm$^2$]      
             & $6.83\times 10^{-52}$ & $5.53\times 10^{-51}$  
             & $5.02\times 10^{-57}$ & $1.34\times 10^{-57}$ & $3.02\times 10^{-55}$ \\
Super-K events    &  6.72 & 54.4  
             & $4.9\times 10^{-5}$ & $1.3\times 10^{-5}$ & $3.0\times 10^{-3}$\\
Hyper-K events    & 163 & 1322  
             & $1.2\times 10^{-3}$ & $3.2\times 10^{-4}$ & $7.2\times 10^{-2}$\\
DUNE events  & 11.0 & 89.4  
             & $8.1\times 10^{-5}$ & $2.2\times 10^{-5}$ & $4.9\times 10^{-3}$\\
\hline
& \multicolumn{2}{c |}{$\bar{\chi} n \to3\pi^0$ \& $5\pi^0$} 
&\multicolumn{3}{c}{
$\bar{\chi} n \to \phi3\pi^0\,(y=0.542)~{\&}~\bar{\chi} n\to \phi5\pi^0\,(y=0.337)$} \\
\hline
$\frac{v}{c}\sigma$ [cm$^2$]    
             & $9.71\times 10^{-47}$  
             & $7.90\times 10^{-46}$ 
             & $2.51\times 10^{-51}$ & $5.42\times 10^{-54}$ & $7.04\times 10^{-50}$ \\
Super-K events    &   $9.59\times 10^{5}$ & $7.78\times 10^{6}$  
             & 24.7 & $5.4\times 10^{-2}$ & 693 \\
Hyper-K events    & $2.32\times 10^{7}$ &  $1.88\times 10^{8}$  
             & 601 & 1.30 & 16824 \\
DUNE events  &  $1.57\times 10^{6}$ & $1.28\times 10^{7}$  
             & 40.7 & $8.8\times 10^{-2}$ & 1137 \\
\hline
\hline
\end{tabular}
\end{adjustbox}
\end{table}

We now calculate the DM-neutron 
annihilation cross section for several possible signals in underground experiments.
First, we determine the value of the mixing angle $\theta$ by requiring the 
neutron dark decay widths to be
$\Delta\Gamma_{n\to \chi \gamma}\simeq 7.1\times 10^{-30}$~GeV
and $\Delta\Gamma_{n\to \tilde{\chi} \gamma}+\Delta\Gamma_{n\to \chi \phi}\simeq 7.1\times 10^{-30}$~GeV
for Models I and~II, respectively.
For Model~II, we also fix $\lambda_\phi=0.04$, 
to obtain the correct DM relic density.

For the benchmark points in Table~\ref{tab-1},
the typical values of the mixing angles are respectively,
$\theta\simeq 10^{-10}$ and $10^{-11}$, in Model~I and Model~II.
In general, $\theta$ in Model~II is one order magnitude 
smaller than that in Model~I,
because 
$$
\frac{\Delta \Gamma_{n\to \chi \gamma}}{\Delta \Gamma_{n\to \chi \phi}}
=\frac{2g^2_n e^2}{|\lambda^2_\phi|}\frac{(1-x_1^2)^3}{\sqrt{f(x_1,x_2)}}
\left(  \frac{m_n-m_{\tilde{\chi}}}{m_n-m_\chi} \right)^2\simeq \mathcal{O}(10^{-2})\,,
$$  
where $f(x_1,x_2)\equiv [(1-x_1)^2-x_2^2][(1+x_1)^2-x_2^2]^3$ 
with $x_1\equiv m_\chi/m_n$ and $x_2\equiv m_\phi/m_n$.

%
 
The diagrams in Fig.~\ref{f1-1} for $\bar{\chi}n\to \gamma \pi^0$  in Model~I yield 
the spin averaged amplitude squared,
\begin{eqnarray}
 \frac{1}{4}\sum |M|^2 &=& |F_{\bar{\chi}\gamma n}(Q^2)|^2\,
  |F_{\bar{\chi}\pi n}(Q^2)|^2  \nonumber \\
  &\times &
  \left\lbrace
   \frac{64\pi\,g^2_{n\pi}}{\Lambda^2_{\chi n}} 
\,\frac{(m^2_\chi-t)\left[(s-m^2_\pi)(m^2_\pi+m^2_n-t)-m^2_\pi (m^2_n-u) \right]}
{(t-m^2_n)^2}
\right\rbrace\,, \nonumber
\end{eqnarray}
in terms of the Mandelstam variables $s\equiv (p_1+p_2)^2=(k_1+k_2)^2$, 
$t\equiv (p_1-k_1)^2=(p_2-k_2)^2$, and $u\equiv (p_1-k_2)^2=(p_2-k_1)^2$.
In the static limit, 
which means the relative velocity of two initial particles approaches zero,
the $\bar\chi$ and $n$ 4-momenta can be respectively approximated by
$(m_\chi,\overrightarrow{0})$ and $(m_n,\overrightarrow{0})$, 
so that
\begin{eqnarray}
 \frac{1}{4}\sum |M|^2 &=& |F_{\bar{\chi}\gamma n}(Q^2)|^2\,
  |F_{\bar{\chi}\pi n}(Q^2)|^2  \nonumber \\
  &\times &
  \left\lbrace
   \frac{64\pi\,g^2_{n\pi}}{\Lambda^2_{\chi n}} 
\,m_n m_\chi (m_n+m_\chi)^2 
\left(
\frac{(m_n+m_\chi)^2-m^2_\pi}{m_n(m_n+m_\chi)^2-m_\chi m^2_\pi}
 \right)
\right\rbrace\,. \nonumber
\end{eqnarray}

The cross section features a $1/v$ behavior in the nonrelativistic limit (applicable for an average DM velocity $v\simeq 10^{-3}c$).
We present values of $\frac{v}{c}\sigma$ in Table~\ref{tab-1}, 
which are independent of the DM velocity. 
This process produces $\gamma$ and $\pi^0$ at about a GeV, which makes
Super-Kamiokande well suited to detect this signal.

We analytically estimate the form factor  suppression for $\bar{\chi}n\to \gamma \pi^0$ 
  as follows.
While the virtual momentum flows in the two diagrams in Fig.~\ref{f1-1} may be
different, in the static limit they are the same:
$-Q^2=(\fr{P}{2}-k_1)^2$.
Therefore, 
the cross section is suppressed by the common factor,
\[
\left(\frac{1-m_n^2/\Lambda_n^2}{1+Q^2/\Lambda_n^2}\right)^{4y}\,.
\]
The maximum suppression ($y=2$) for a typical value of $Q^2$ 
is $\mathcal{O}(10^{-5})$.
Note that the form factor suppression for  $\bar{\chi}n\to \phi\gamma \pi^0$  in Model~II
depends on the kinematics and a full numerical integration is required.
The values of $\frac{v}{c}\sigma$ including form factor suppression with $y=2$ are provided in Table~\ref{tab-1}.

\begin{figure}[t!]
\centering
\includegraphics[height=2.9in,angle=270]{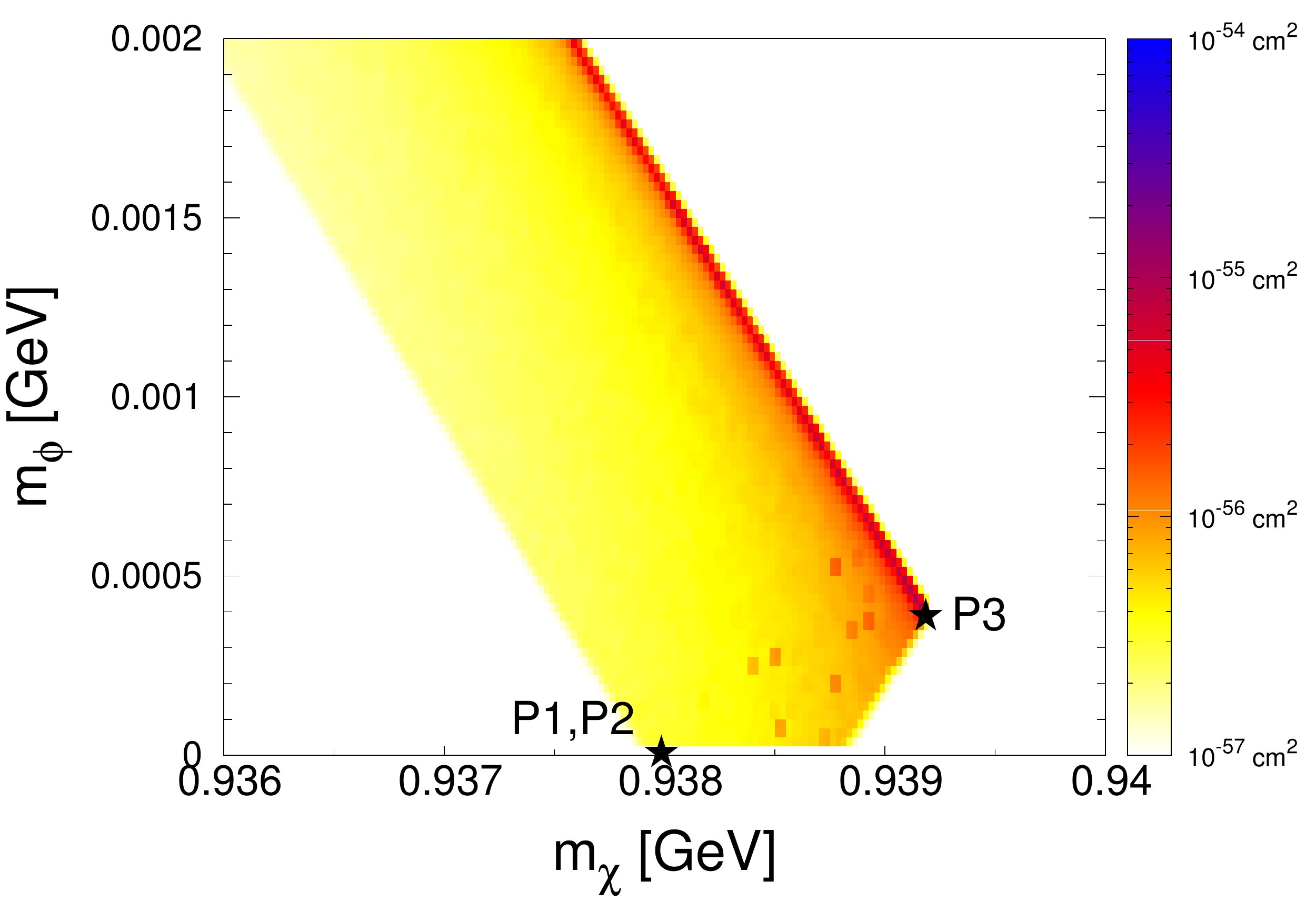}
\includegraphics[height=2.9in,angle=270]{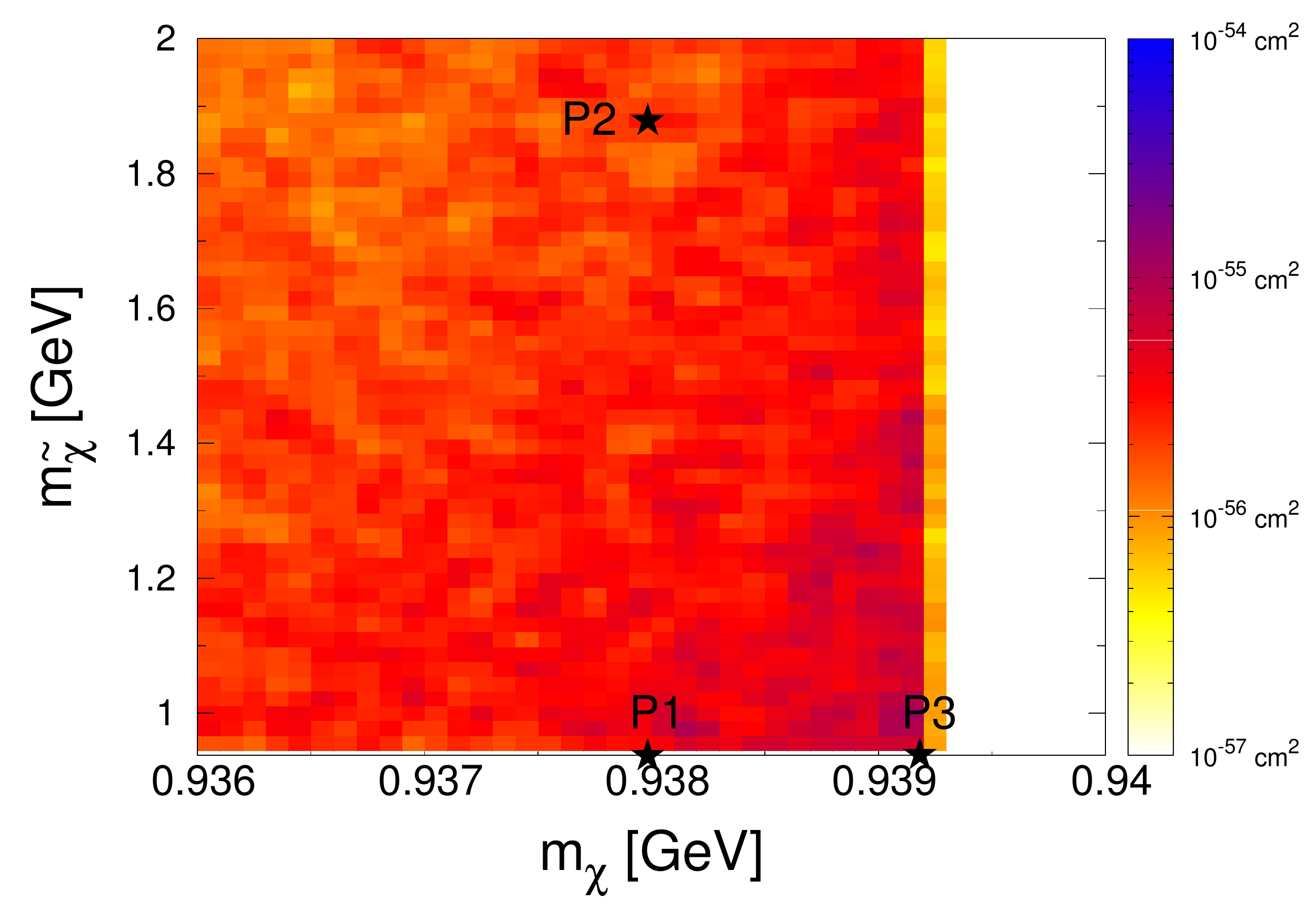}
\includegraphics[height=2.9in,angle=270]{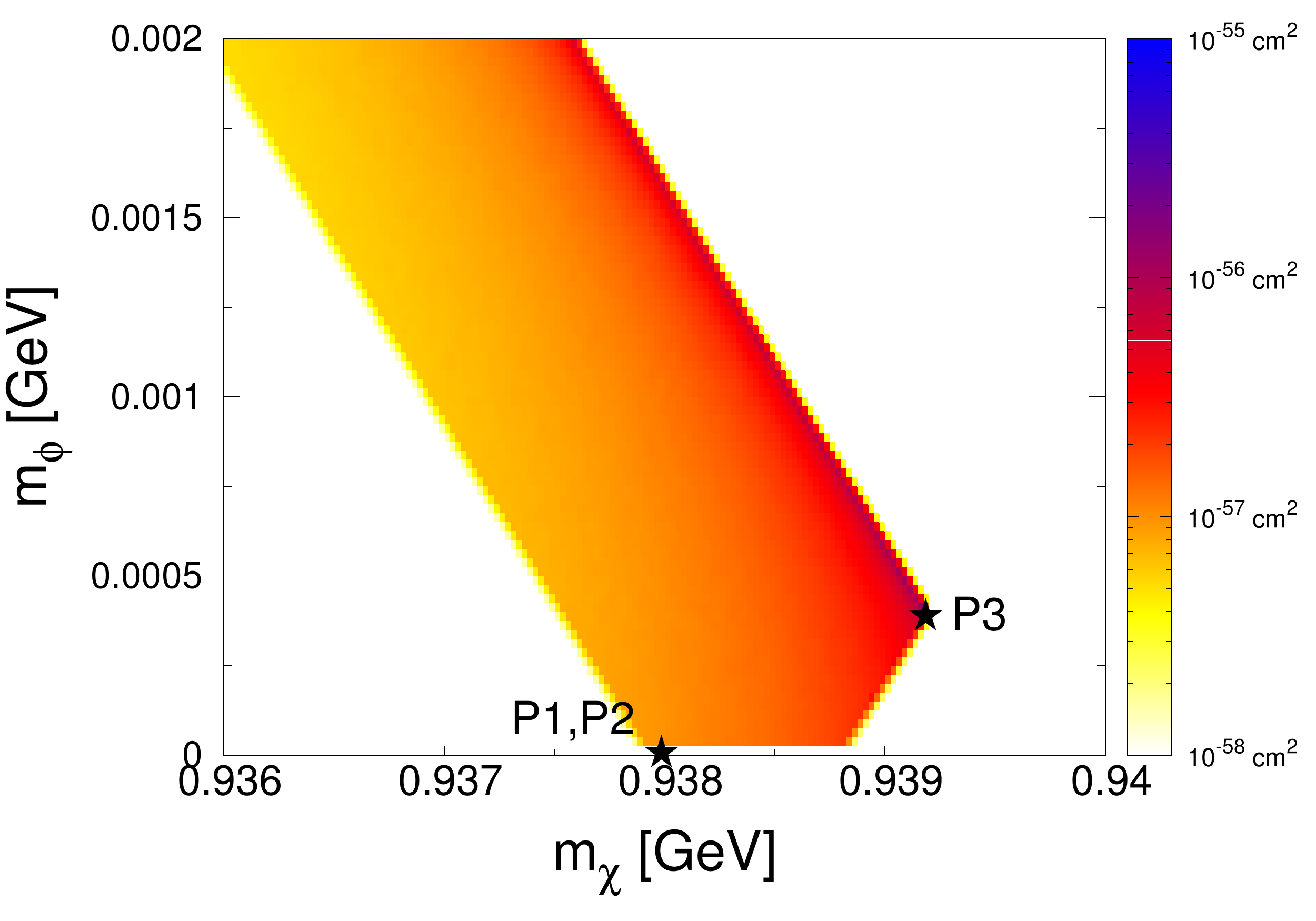}
\includegraphics[height=2.9in,angle=270]{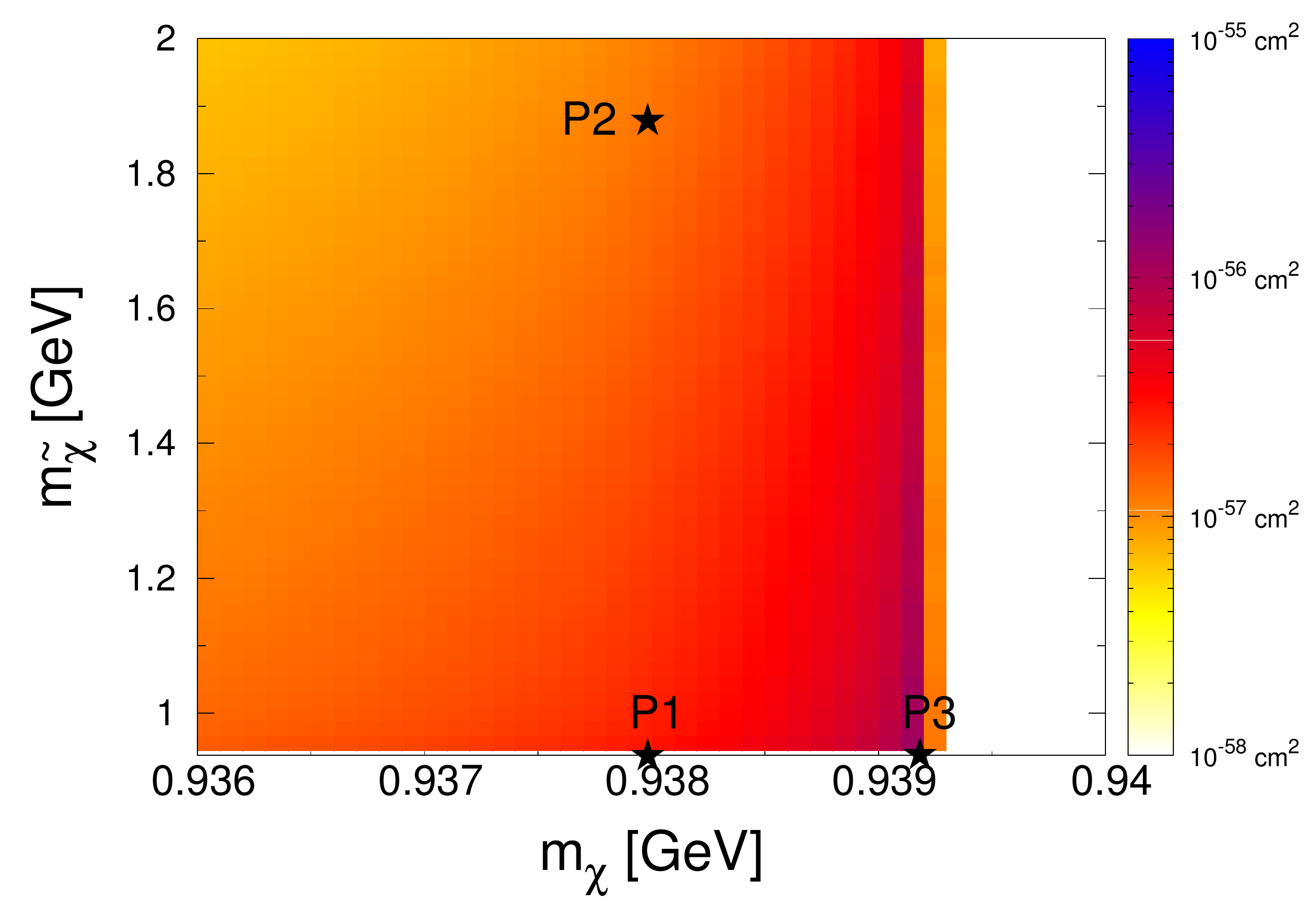}
\caption{\small \label{fig-sigma} 
The maximum (upper panels), and minimum (lower panels) values of $\frac{v}{c}\sigma(\bar{\chi} n \to \gamma\pi^0\phi)$ projected on the $(m_\chi$,$m_\phi)$ plane (left panels), and the $(m_\chi,m_{\tilde{\chi}})$ plane (right panels). 
The stars mark the three benchmark points {\bf P1}, {\bf P2}, and {\bf P3}.  We set $y=2$. 
}
\end{figure}

Because Model~II has multiple free parameters, $(m_\chi,m_\phi,m_{\tilde{\chi}})$, 
 the 
$\bar{\chi}n \to \phi\gamma\pi^0$ cross section is enhanced in some regions of parameter space.
The maximum value of 
$\frac{v}{c}\sigma\simeq \mathcal{O}(10^{-55})$~cm$^2$ occurs
at the corner of the parameter space, 
when three conditions  are satisfied:
i) $m_\chi+m_\phi \to 939.565~{\rm MeV}$,
ii) $|m_\chi-m_\phi|\to m_p+m_e$,
and iii) $m_{\tilde{\chi}} \to 937.992~{\rm MeV}$.
{\bf P3} is such a benchmark point with 
$\sigma(\bar{\chi}n \to \phi\gamma\pi^0)=3.02\times 10^{-55}$~cm$^2$.
The distribution of maximum and minimum cross section values in the $(m_\chi,m_\phi,m_{\tilde{\chi}})$ parameter space are shown in 
Fig.~\ref{fig-sigma}. The benchmark points
 {\bf P1}, {\bf P2}, and {\bf P3} are marked with stars.

We now compute the DM-neutron annihilation cross section to 
multi-$\pi^0$.
For Model~I, the process is $\bar{\chi}n\to{\rm multi}$-$\pi^0$, and
for Model~II, $\phi$ is produced in association with multi-$\pi^0$,
$\bar{\chi}n\to\phi+$multi-$\pi^0$.
Since the multi-pion channel is the dominant mode for 
antinucleus-nucleus annihilation,
we expect the same for DM-neutron annihilation.

Since a perturbation calculation is not valid for a large $\bar{n}\pi n$ coupling, $g_{n\pi}/\sqrt{4\pi}\simeq \mathcal{O}(1)$,
we use the experimentally measured value of the $\bar{n}p$ annihilation cross section
in the low $\bar{n}$ velocity limit~\cite{Mutchler:1988av,Feliciello:1999ti,Bertin:1997gn,Armstrong:1987nu}, obtained by extrapolating the
cross section in Ref.~\cite{Mutchler:1988av} to zero momentum:
$$
v\sigma(\bar{n}p \to {\text{multi-pions}})_{\rm exp}=44\pm 3.5~ \rm mb\,, \nonumber
$$
which is s-wave dominant and independent of the $\bar{n}$ velocity.
We assume that $\sigma(\bar{n}n)\simeq \sigma(\bar{n}p)$.
Then for Model~I, the DM-neutron annihilation cross section is given by
$$
v\sigma(\bar{\chi}n\to {\text{multi-pions}})
=\theta^2\,v\sigma(\bar{n}p \to {\text{multi-pions}})_{\rm exp}\,.
$$
Numerical values for the sum of the cross section to the $3\pi^0$ and $5\pi^0$ channels, which have a total branching ratio of 
0.585~\cite{Abe:2011ky}, are provided in Table~\ref{tab-1}.

For $\bar{\chi}n \to\phi+$multi-$\pi^0$ in Model~II,
there is no experimental dataset that can be used directly.
Our strategy is to perturbatively calculate
$\bar{n}n\to {\rm pions}$ cross sections and then
require these to match Super-K's simulated cross sections
in Table~I of Ref.~\cite{Abe:2011ky} by tuning the exponent $y$ of the form factor.
We find $y=0.542$ and $y=0.337$ 
for $\bar{n}n\to 3\pi^0$ and $\bar{n}n\to 5\pi^0$, respectively; see the Appendix for a description of our procedure.
Using these  values of $y$ in Eq.~(\ref{eq:npin}) 
and the mixing angle $\theta$,
we calculate the cross sections for 
$\bar{\chi}n\to \phi3\pi^0$ and $\bar{\chi}n\to \phi5\pi^0$ as outlined in the Appendix. 
Values for the sum of the cross section to these two channels are given in Table~\ref{tab-1}.


\section{Signal events at Super-K, Hyper-K and DUNE}
\label{sec:signal}

Armed with the $\bar{\chi}n$ 
annihilation cross section for different channels,
the DM-nucleus annhilation cross section can be determined from $\sigma(\bar{\chi}A)=A^{2/3}\sigma_0$~\cite{Astrua:2002zg}, 
where $A$ is the atomic mass of the nucleus of atomic number $Z$, and
$\sigma_0=\alpha \sigma(\bar{\chi}p)+(1-\alpha)\sigma(\bar{\chi}n)$,
with $\alpha\equiv Z/A$. 
In the following, we make the assumption that 
$\sigma(\bar{\chi}p)=0$.

With water as the target for Super-K  and Hyper-K, and $m_\chi=938.783$ MeV, 
the interaction rate per second per gram of water is
$$
n_\chi v_{\rm DM}[(N_A\cdot 1/18)\cdot \sigma(\bar{\chi}O)+
                  (N_A\cdot 2/18)\cdot \sigma(\bar{\chi}H)]\,,
$$
where $v_{\rm DM}=10^{-3}c$ is the thermal average DM velocity, the DM number density is  $n_{\chi}=\rho_{\chi}/m_\chi$ per cm$^3$ in terms of the local DM density  $\rho_{\chi}=0.3~{\rm GeV/cm^3}$, $N_A=6.022\times 10^{23}$ is the Avogadro number, 
and $N_A/18$ is the total number of 
${\rm H_2O}$ molecules per gram of water. 
For the liquid Argon target at DUNE, the interaction rate per second per gram of target is
$$
n_\chi v_{\rm DM}[(N_A/40)\cdot \sigma(\bar{\chi}Ar)]/\rho_{\rm Ar}\,,
$$
where $\rho_{\rm Ar}=1.3954\,{\rm g/cm^3}$ is the density of liquid Argon.
In the nonrelativistic limit, 
the interaction rate is independent of the $v_{\rm DM}$, and
therefore independent of the velocity-distribution of the DM in the galactic halo. 

The signal events are obtained by multiplying the above interaction rates
with the total exposure.
For Super-K~\cite{Miura:2016krn}, 
the current total exposure is 306.3 kiloton-years.
For Hyper-K~\cite{Abe:2018uyc},
we use a fiducial mass 
of 372 kiloton with 20 years of data-taking. 
For DUNE~\cite{Acciarri:2015uup}, we take a 40 kiloton fiducial mass with 20 years of data-taking.
The events numbers for the different signal channels in the three experiments are displayed in Table~\ref{tab-1}.

The kinematic cuts applied in Super-K's searches for proton decay and $n-\bar{n}$ oscillations
are summarized in Table~\ref{tab:cut}.
We adopt the same cuts ({\bf cut-1}, {\bf cut-2}, {\bf cut-3})
and definitions of total visible momentum,
$P_{\rm tot}\equiv |\sum^{\rm all-rings}_i \overrightarrow{p_i}|$,
where $\overrightarrow{p_i}$ is the reconstructed momentum vector of the $i^{\rm th}$ ring,
the invariant mass, $M_{\rm tot}\equiv \sqrt{E^2_{\rm tot}-P^2_{\rm tot}}$,
and the total visible energy, $E_{\rm tot}\equiv \sum^{\rm all-rings}_i\sqrt{p^2_i+m^2_i}$, 
where $m_i$ is the mass of the $i^{\rm th}$ ring assuming that showering 
and nonshowering rings are from $\gamma$ and $\pi^\pm$,
respectively~\cite{Abe:2011ky}.
For our case, $m_i=0$.
Kinematic {\bf cut-1} was applied for the $n-\bar{n}$ oscillation search
for which  the observed number of events $N_{\rm obs}=24$ is consistent 
with the number of background events $N_{\rm bkgd}=24.1$.
Correspondingly, the $3\sigma$ range of the allowed number of signal events is
$N^{3\sigma}_{\text {Super-K}}\subset [0,22.5]$~\cite{Feldman:1997qc};
the allowed number of signal events
for {\bf cut-2} and {\bf cut-3} are as in Table~\ref{tab:cut}.
To evaluate the expected number of signal events at Hyper-K and DUNE, 
we assume that the observed event rate is compatible 
with the expected background rate, and
scale Super-K's exposure.
The $3\sigma$ ranges are provided in Table~\ref{tab:cut}.

\begin{table}[t]
\caption{\small \label{tab:cut}
Three kinematic regions from $n-\bar{n}$ oscillations~\cite{Abe:2011ky}
and proton decay~\cite{Miura:2016krn} searches at Super-K. 
$N^{3\sigma}_{\text {Super-K}}$ is the allowed number of signal events within $3\sigma$.
The $3\sigma$ expectation for the number of signal events at Hyper-K and DUNE is obtained under 
the assumption that the observed number of events is compatible with the number of background events.
}
\begin{adjustbox}{width=\textwidth}
\begin{tabular}{c|c|ccc|c|c}
\hline
\hline
    & Kinematic cuts (in MeV) & $N_{\rm obs}$ & $N_{\rm bkgd}$ & $N^{3\sigma}_{\text {Super-K}}$ 
    & $N^{3\sigma}_{\text {Hyper-K}}$ & $N^{3\sigma}_{\rm DUNE}$ \\
\hline
{\bf cut-1}    & $P_{\rm tot}\subset [0,450]~M_{\rm tot}\subset[750,1800]$~\cite{Abe:2011ky} 
               & 24  & 24.1  & $[0,22.5]$ 
               &  $[0,75]$   & $[0,27]$ \\
\hline
{\bf cut-2}    & $P_{\rm tot}\subset [0,100],~M_{\rm tot}\subset[800,1050]$~\cite{Miura:2016krn} 
               & 0  & 0.07  & $[0,7]$
               & $[0,5.5]$ & $[0,4]$  \\
\hline
{\bf cut-3}    & $P_{\rm tot}\subset [100,250],~M_{\rm tot}\subset[800,1050]$~\cite{Miura:2016krn} 
               & 0  & 0.54  & $[0,6.5]$
               & $[0,7]$  & $[0,5.8]$  \\
\hline
\hline
\end{tabular}
\end{adjustbox}
\end{table}

\begin{figure}[t!]
\centering
\includegraphics[height=2.9in,angle=270]{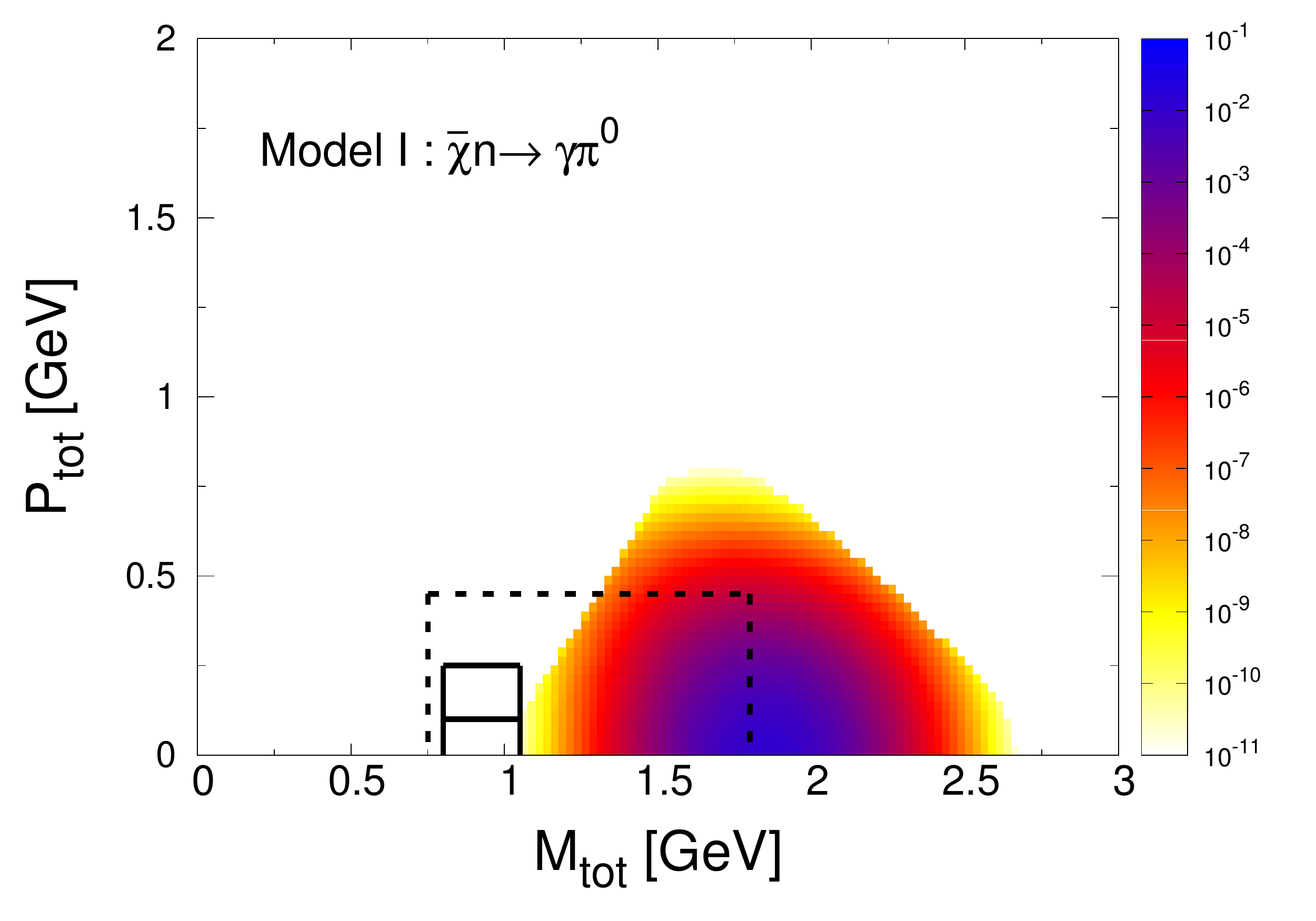}
\includegraphics[height=2.9in,angle=270]{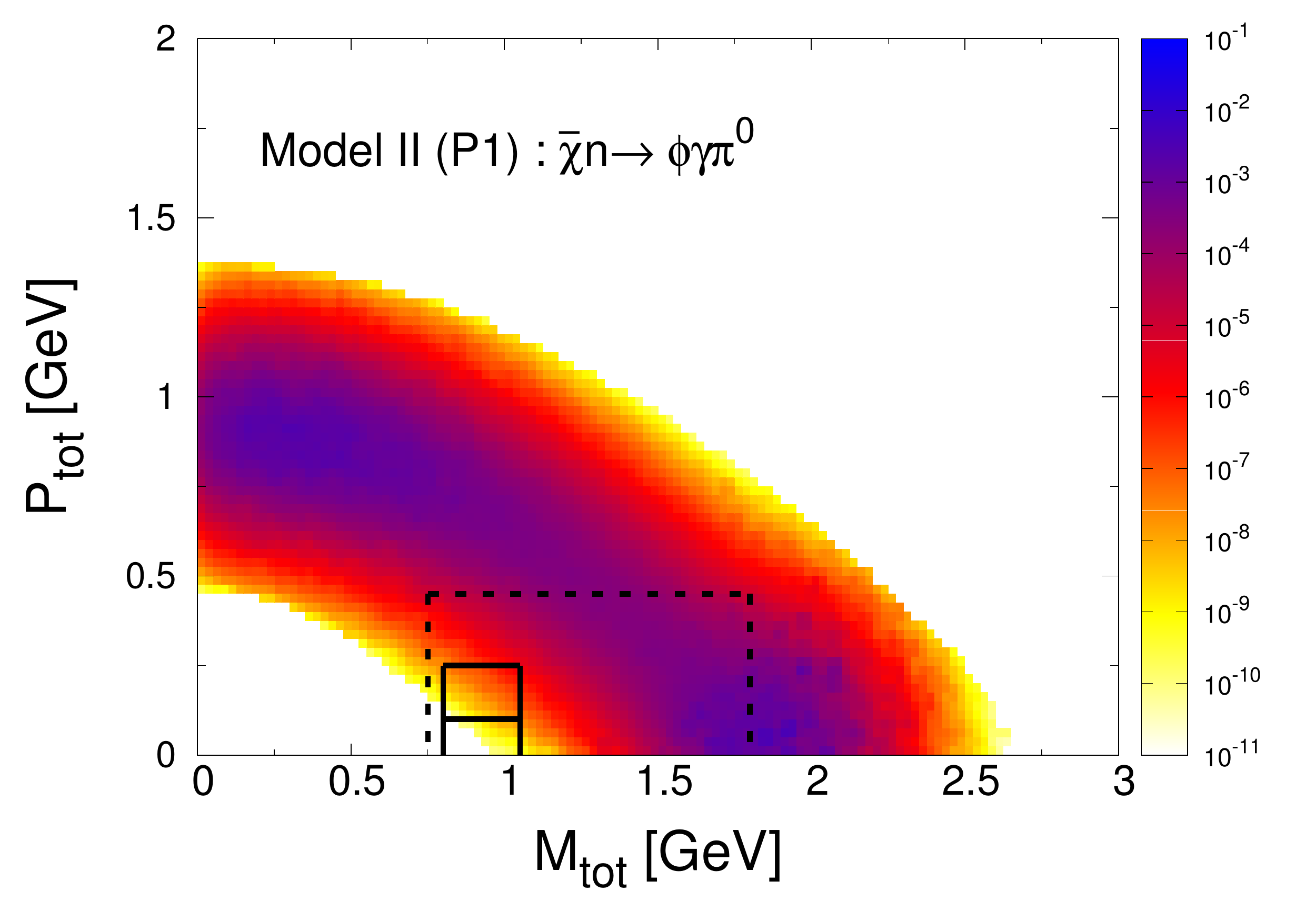}
\includegraphics[height=2.9in,angle=270]{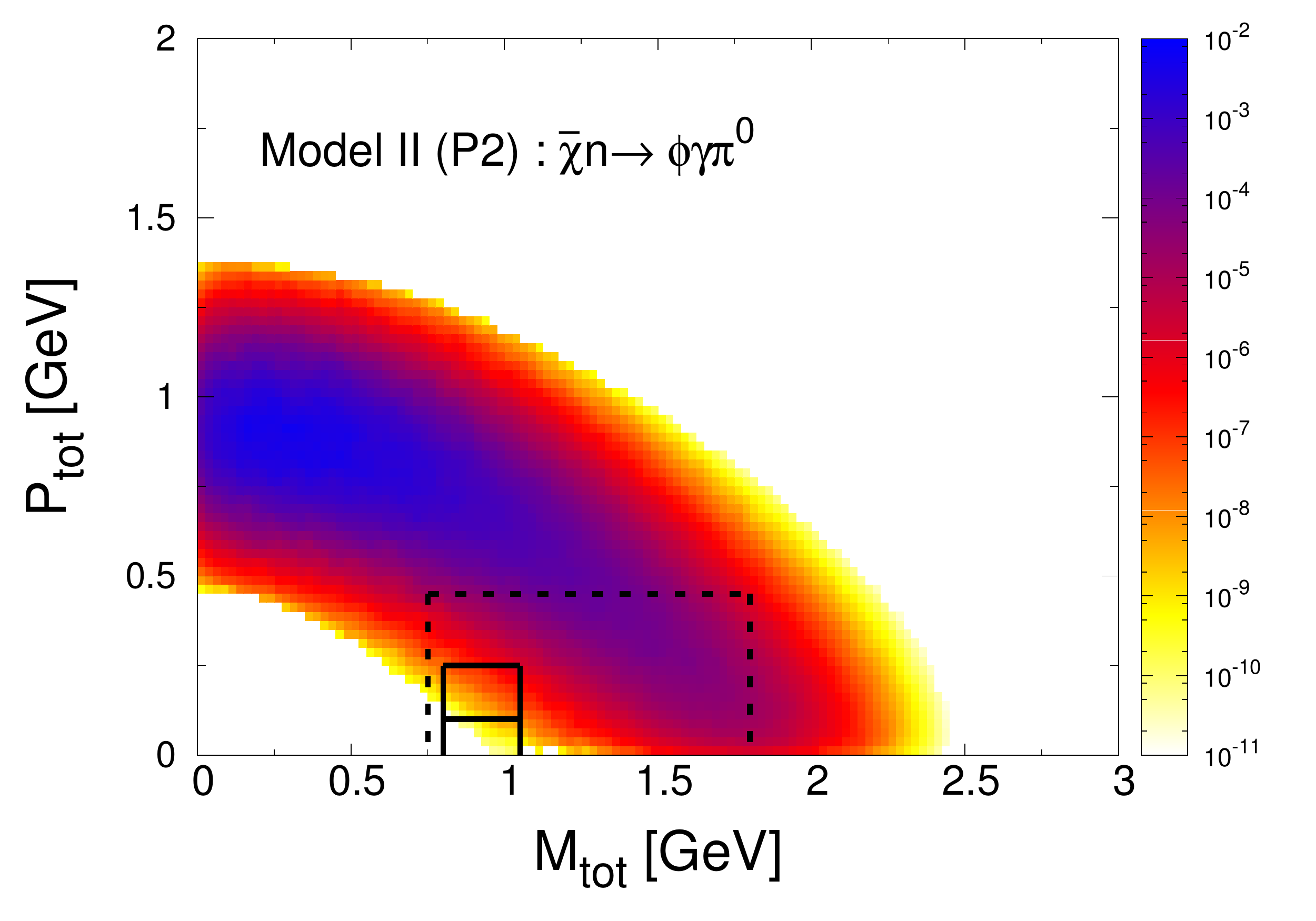}
\includegraphics[height=2.9in,angle=270]{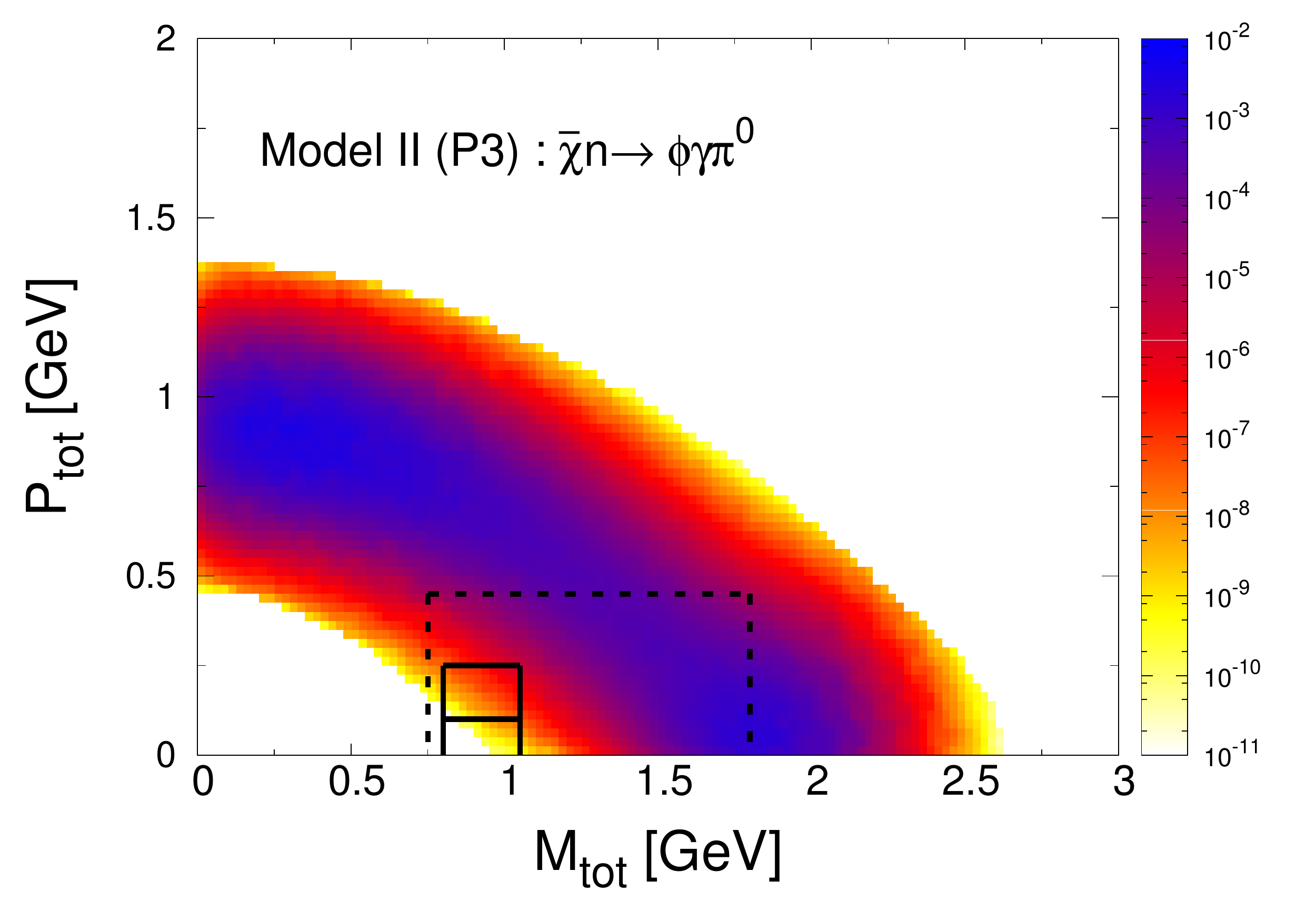}
\caption{\small \label{fig:api} 
The normalized signal event distributions for  $\bar{\chi}n\to \gamma \pi^0$ and $\bar\chi n \to \phi \gamma \pi^0$
 in the $(M_{\rm tot},P_{\rm tot})$ plane for Model~I (with $m_\chi=937.992$~MeV) and the three benchmark points of Model~II. The dashed rectangle corresponds to kinematic {\bf cut-1}, 
while two solid lower and upper rectangles correspond to {\bf cut-2} and {\bf cut-3}, respectively. 
}
\end{figure}

\begin{figure}[t!]
\centering
\includegraphics[height=2.9in,angle=270]{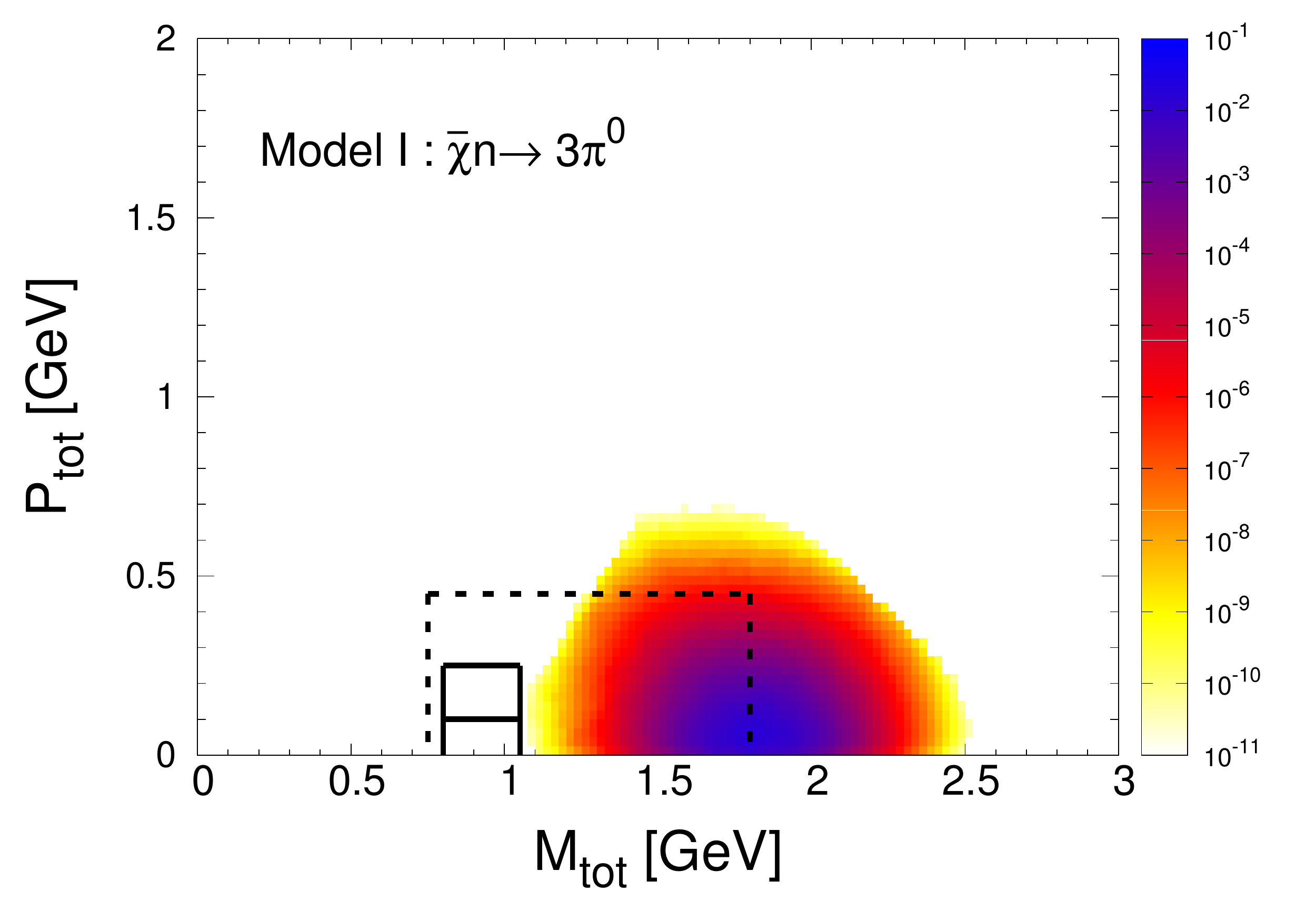}
\includegraphics[height=2.9in,angle=270]{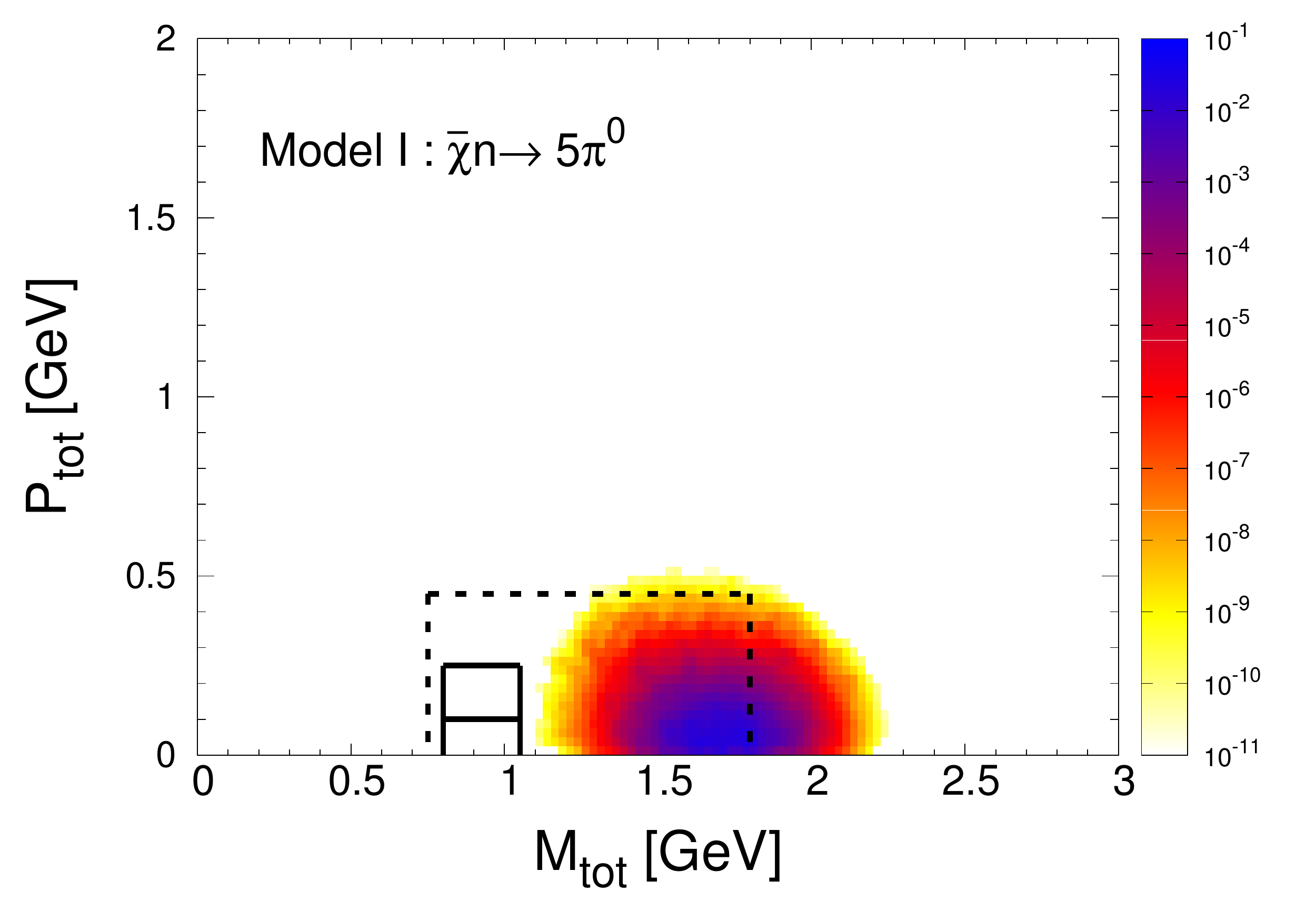}
\includegraphics[height=2.9in,angle=270]{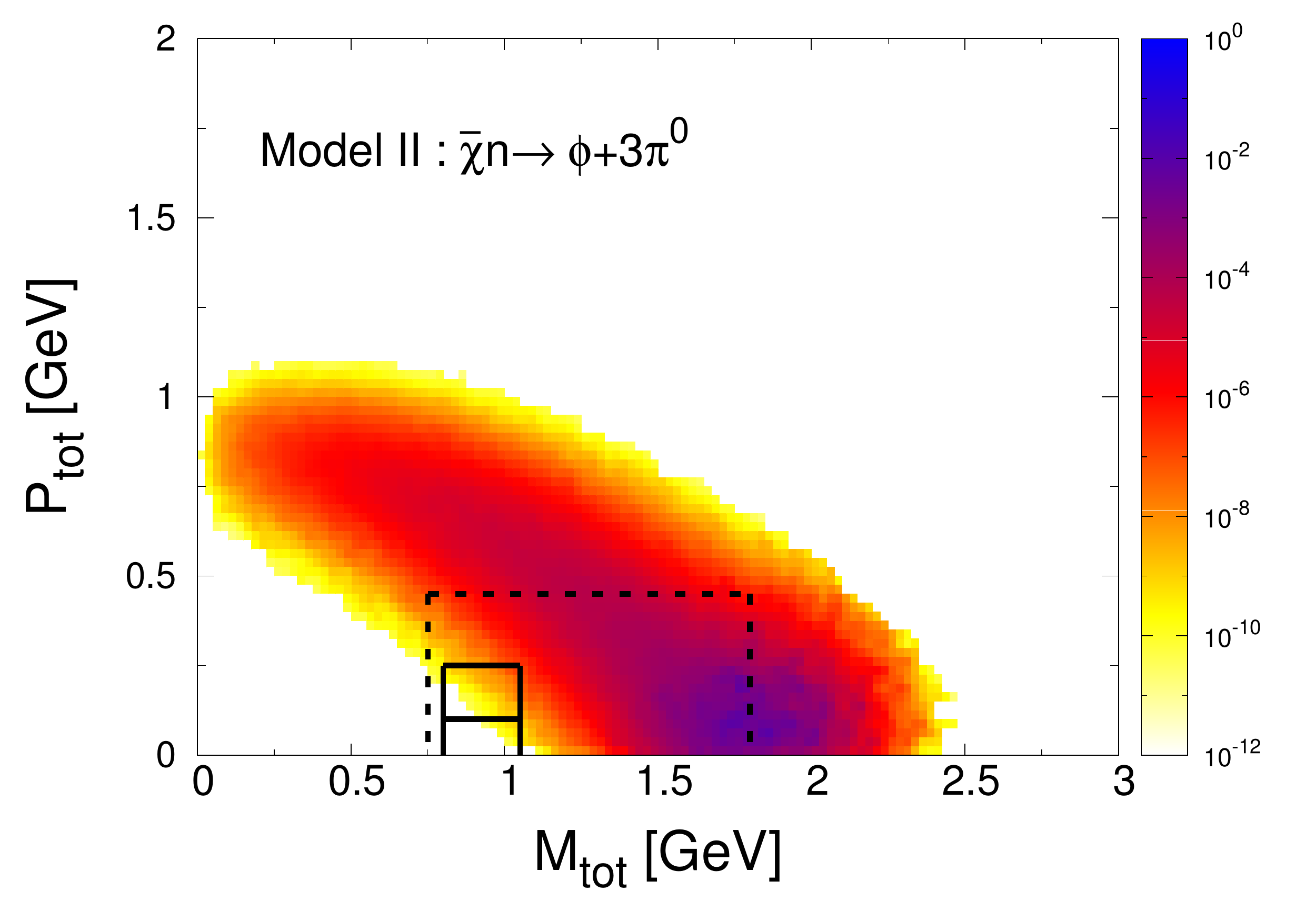}
\includegraphics[height=2.9in,angle=270]{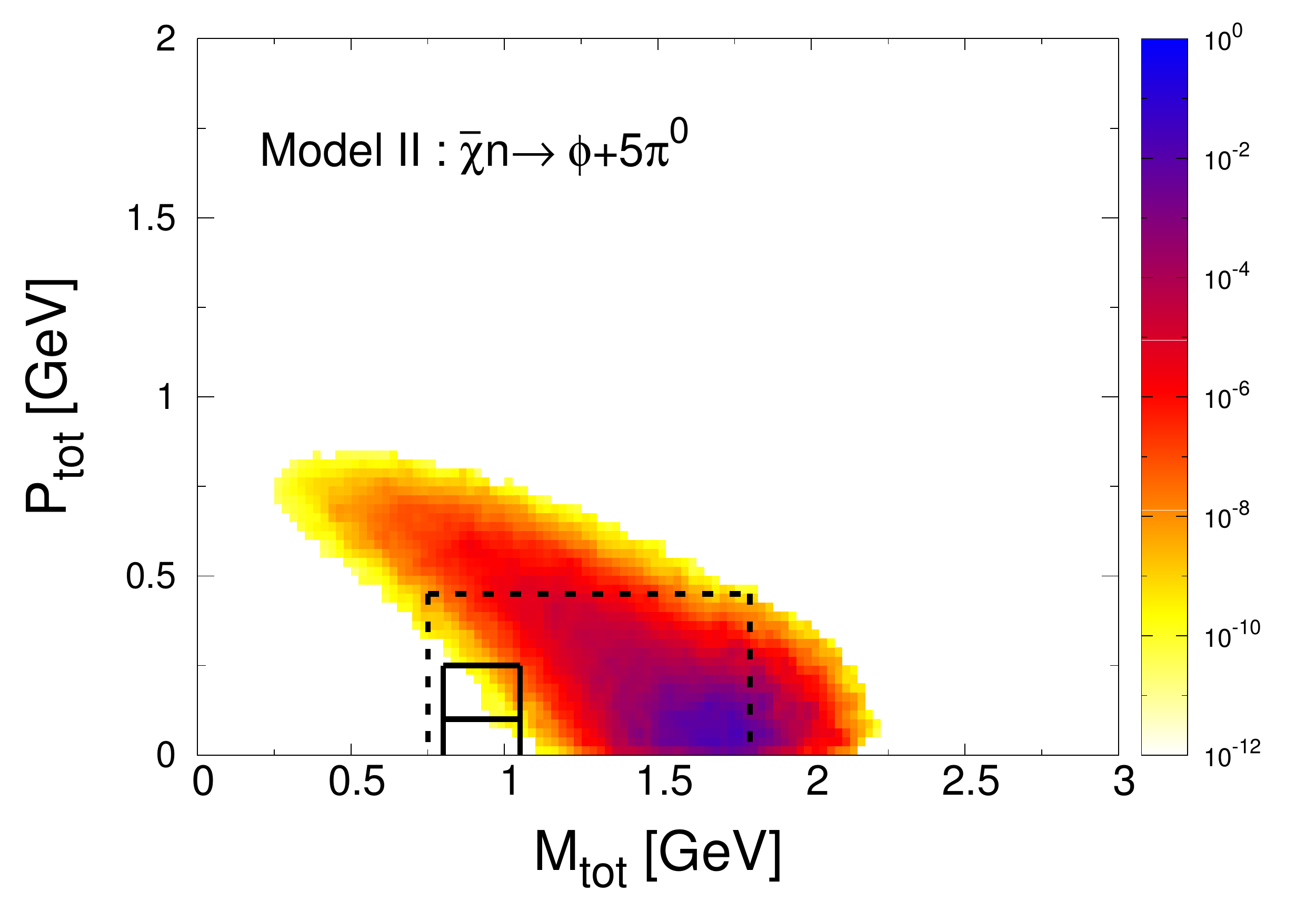}
\caption{\small \label{fig:pions} 
The normalized signal event distributions 
for $\bar{\chi}n\to 3\pi^0~(5\pi^0)$ for Model~I (with $m_\chi=937.992$~MeV) and $\bar{\chi}n\to \phi3\pi^0~(\phi5\pi^0)$ for point {\bf P1} of Model~II .
}
\end{figure}

To calculate the number of events that satisfy the kinematic cuts,
we perform a Monte Carlo simulation by assuming
10\% momentum uncertainty for each ring in Super-K~\cite{Ashie:2005ik}.\footnote{The momentum resolution is estimated to be 
$0.6+2.6\sqrt{\rm P(GeV/c)}\%$ for Super-K~\cite{Ashie:2005ik}.
Since for our signal processes, each ring has about a few hundred~MeV in energy, 
we simply adopt a 10\% momentum resolution.} We take the momentum resolution at Hyper-K and DUNE to be 10\%.
The event distributions projected on to the $(M_{\rm tot},P_{\rm tot})$ plane for
$\bar{\chi}n\to \gamma \pi^0\,(\phi \gamma \pi^0)$ and
$\bar{\chi}n\to {\text {multi-}}\pi^0\,(\phi+$multi-$\pi^0$) are shown in Figs.~\ref{fig:api} and~\ref{fig:pions}, respectively. 
Kinematic {\bf cut-1} is the region within the dashed rectangle, 
and {\bf cut-2} ({\bf cut-3}) is within the solid lower (upper) rectangle.
In Fig.~\ref{fig:api}, we show the $\bar{\chi}n\to \gamma \pi^0$ event distribution for the $m_\chi=937.992$~MeV case of Model~I, but the distribution is not visibly changed for the $m_\chi=938.783$~MeV case.
 The $\bar{\chi}n \to \phi \gamma \pi^0$ event distributions are shown
for points {\bf P1}, {\bf P2}, and {\bf P3} of Model II;
less then 20\% of the total events fall in the {\bf cut-1} region.
Because the kinematic distributions of our multi-$\pi^0$ signals are similar to that of 
$n-\bar{n}$ oscillations, in Fig.~\ref{fig:pions}, a majority of the events fall inside the {\bf cut-1} region, 
and only a tiny fraction of events are inside the {\bf cut-2} and {\bf cut-3} regions.
Therefore, these signals do not contaminate the proton decay search.
In Table~\ref{tab:pass}, we tabulate the percentage of events for each channel that pass
the three kinematic cuts.

\begin{table}[t]
\caption{\small \label{tab:pass}
Percentage of events that pass the kinematic cuts.
}
\begin{adjustbox}{width=\textwidth}
\begin{tabular}{c|ccc|ccc}
\hline
\hline
 & \multicolumn{3}{c |}{{\bf Model~I: $m_\chi=937.992$}~MeV}  & \multicolumn{3}{c}{{\bf Model~II: P1}}  \\
\hline
 & $\bar{\chi} n \to \gamma \pi^0$ & $\bar{\chi} n \to 3\pi^0$ & $\bar{\chi} n \to 5\pi^0$ 
 & $\bar{\chi} n \to \phi \gamma \pi^0$ & $\bar{\chi} n \to \phi3\pi^0$ & $\bar{\chi} n \to \phi5\pi^0$   \\
\hline
{\bf cut-1}    & $31.2~\%$ & $41.0~\%$ & $79.7~\%$ 
               & $15.4~\%$ & $78.3~\%$ & $71.8~\%$  \\
{\bf cut-2}    & $2.9\times 10^{-9}~\%$ & $1.1\times 10^{-9}~\%$  & $5.7\times 10^{-9}~\%$ 
               & $2.4\times 10^{-6}~\%$ & $2.4\times 10^{-7}~\%$ & $1.5\times 10^{-7}~\%$   \\
{\bf cut-3}    & $2.7\times 10^{-10}~\%$ & $5.7\times 10^{-10}~\%$ & $1.0\times 10^{-10}~\%$ 
               & $3.8\times 10^{-4}~\%$ & $2.3\times 10^{-5}~\%$ & $1.5\times 10^{-5}~\%$ \\
\hline
 & \multicolumn{3}{c |}{{\bf Model~II: P2}} & \multicolumn{3}{c}{{\bf Model~II: P3}}  \\
\hline
 & $\bar{\chi} n \to \phi\gamma \pi^0$ & $\bar{\chi} n \to \phi3\pi^0$ & $\bar{\chi} n \to \phi5\pi^0$ 
 & $\bar{\chi} n \to \phi \gamma \pi^0$ & $\bar{\chi} n \to \phi3\pi^0$ & $\bar{\chi} n \to \phi5\pi^0$   \\
\hline
{\bf cut-1}    & $1.76~\%$ & $57.6~\%$ 
               & $93.5~\%$ 
               & $14.6~\%$ & $57.5~\%$ 
               & $87.3~\%$  \\
{\bf cut-2}    & $1.3\times 10^{-6}~\%$ & $7.8\times 10^{-6}~\%$  
               & $4.6\times 10^{-6}~\%$ 
               & $3.2\times 10^{-6}~\%$ & $1.0\times 10^{-6}~\%$  
               & $3.6\times 10^{-7}~\%$   \\
{\bf cut-3}    & $2.8\times 10^{-4}~\%$ & $1.1\times 10^{-3}~\%$ 
               & $1.1\times 10^{-3}~\%$
               & $5.0\times 10^{-4}~\%$ & $1.0\times 10^{-4}~\%$  
               & $5.8\times 10^{-5}~\%$ \\
\hline
\hline
\end{tabular}
\end{adjustbox}
\end{table}

\section{Results and summary}
\label{sec:result}

Under the assumption that $\bar{\chi}$ is stable on the scale of the age of universe and is the dominant component of dark matter,
Model~I is comfortably ruled out by the current $n-\bar{n}$ oscillation search at Super-K because it predicts $\mathcal{O}(10^6)$ 
$\bar{\chi}n \to {\rm multi}$-$\pi^0$ events in the {\bf cut-1} region, while Super-K has observed 24 events with an expected background of 24.1 events.
Note that theoretical uncertainties do not affect this exclusion because the calculation of the $\bar{\chi}n \to {\rm multi}$-$\pi^0$ cross section is driven by experimental data, and so is not impacted by
the hadron form factor uncertainty.

It is difficult to completely explore the parameter space of Model~II because of its many degrees of freedom, and hence difficult to rule it out. We therefore focused on specific benchmark points.
The expected numbers of $\phi3\pi^0+\phi5\pi^0$ signal events for {\bf P1}, {\bf P2}, and {\bf P3} at Super-K
after applying {\bf cut-1}
are 18.1, 0.040, and 545, respectively,
where we used $y=0.542$ and $y=0.337$  for 
$\bar{\chi}n \to \phi 3\pi^0$ and $\bar{\chi}n \to \phi 5\pi^0$, respectively.
It is clear that {\bf P3} is excluded by Super-K at more than $3\sigma$ for the above values of $y$. 
{\bf P1} does not contribute a significant event excess at Super-K. 
{\bf P2} is three orders of magnitude beyond the reach of Super-K because of the heavier $\tilde{\chi}$.

If {\bf cut-1} is extended to $M_{\rm tot}=2$~GeV,
the signal events increase to 24.4, 0.041, and 680 for {\bf P1}, {\bf P2}, and {\bf P3}, respectively,
and more than 95\% of the signal events fall inside the extended kinematic region for {\bf P1} and {\bf P3}.

We show the sensitivities 
of  Super-K and the future experiments Hyper-K and DUNE in terms of  $y$
in Table~\ref{tab-y}, where we applied kinematic {\bf cut-1}. The table gives the minimum
value of $y$ that ensures that the number of signal events lies within the $3\sigma$ range in Table~\ref{tab:cut}.
Negative values of $y$ mean that although there is no form factor suppression,
the experiment cannot probe the parameter point.
%
Clearly, DUNE will have better sensitivity than Super-K, 
and Hyper-K will have the best sensitivity as evidenced by the higher minimum values of $y$.
%

  \begin{table}[t]
\caption{\small \label{tab-y}
The minimum value of $y$ for Model~II that produces
a signal event number within the $3\sigma$ range in Table~\ref{tab:cut}; the maximum value (which gives 0 events) is $y=\infty$.
Here kinematic {\bf cut-1} is applied.
}
\begin{tabularx}{\textwidth}{c|@{\hskip 1in}c@{\hskip 1in}c@{\hskip 1in}c@{\hskip 1in}}
\hline
\hline
 \multicolumn{4}{c}{Super-K}  \\
\hline
    & {\bf P1} & {\bf P2}  & {\bf P3}  \\
\hline
$\bar{\chi}n \to \phi \gamma \pi^0$      
             &  -0.807 & -3.48 & -0.236  \\
\hline
$\bar{\chi}n \to \phi 3\pi^0$      
             & 0.229 & -0.721 & 0.883  \\
\hline
$\bar{\chi}n \to \phi 5\pi^0$      
             & 0.260 & -0.502 & 0.735  \\
\hline
\hline
 \multicolumn{4}{c}{Hyper-K}  \\
\hline
   & {\bf P1} & {\bf P2}  & {\bf P3}  \\
\hline
$\bar{\chi}n \to \phi \gamma \pi^0$      
             & -0.434 & -2.88 & 0.172  \\
\hline
$\bar{\chi}n \to \phi 3\pi^0$      
             & 0.658 & -0.371 & 1.297  \\
\hline
$\bar{\chi}n \to \phi 5\pi^0$      
             & 0.535 & -0.261 & 1.003  \\
\hline
\hline
 \multicolumn{4}{c}{DUNE}  \\
\hline
    & {\bf P1} & {\bf P2}  & {\bf P3}  \\
\hline
$\bar{\chi}n \to \phi \gamma \pi^0$      
             & -0.751 & -3.38 & -0.173  \\
\hline
$\bar{\chi}n \to \phi 3\pi^0$      
             & 0.296 & -0.665& 0.948  \\
\hline
$\bar{\chi}n \to \phi 5\pi^0$      
             & 0.304 & -0.464 & 0.777 \\
\hline
\hline
\end{tabularx}
\end{table}

\section*{Acknowledgments}  
W.-Y.K. and P.-Y.T. thank the National Center of Theoretical Sciences, 
Taiwan, for its hospitality. D.M. is supported in
part by the U.S. DOE under Grant No. de-sc0010504.
P.-Y.T. is supported by World Premier International Research 
Center Initiative (WPI), MEXT, Japan. 
\newpage
\appendix

\section*{Appendix}
\label{app:a}

We describe the procedure to obtain the 
$\bar{\chi}n \to \phi 3\pi^0$ and $\bar{\chi}n \to \phi 5\pi$ 
cross sections including the form factor suppressions.

From a theoretical perspective, it is difficult to calculate 
the antineutron-neutron cross section and branching ratios
because non-perturbative pion interactions are involved.
Instead, we extract the annihilation cross sections and branching ratios
from relevant experimental data, 
and then use them to determine the unknown exponent $y$ in the form factor of Eq.~(\ref{eq:ff}).
Higher-order corrections, pion self interactions, 
and form factor uncertainties are embedded 
in this form factor. 
The values of $y$ can be different for different final states because our  ``effective form factor''  includes the aforementioned effects.
We emphasize that our form factor is {\it not} one that underlies the fundamental process that gives different final states. 

\begin{figure}[t!]
\centering
\includegraphics[height=2.9in,angle=270]{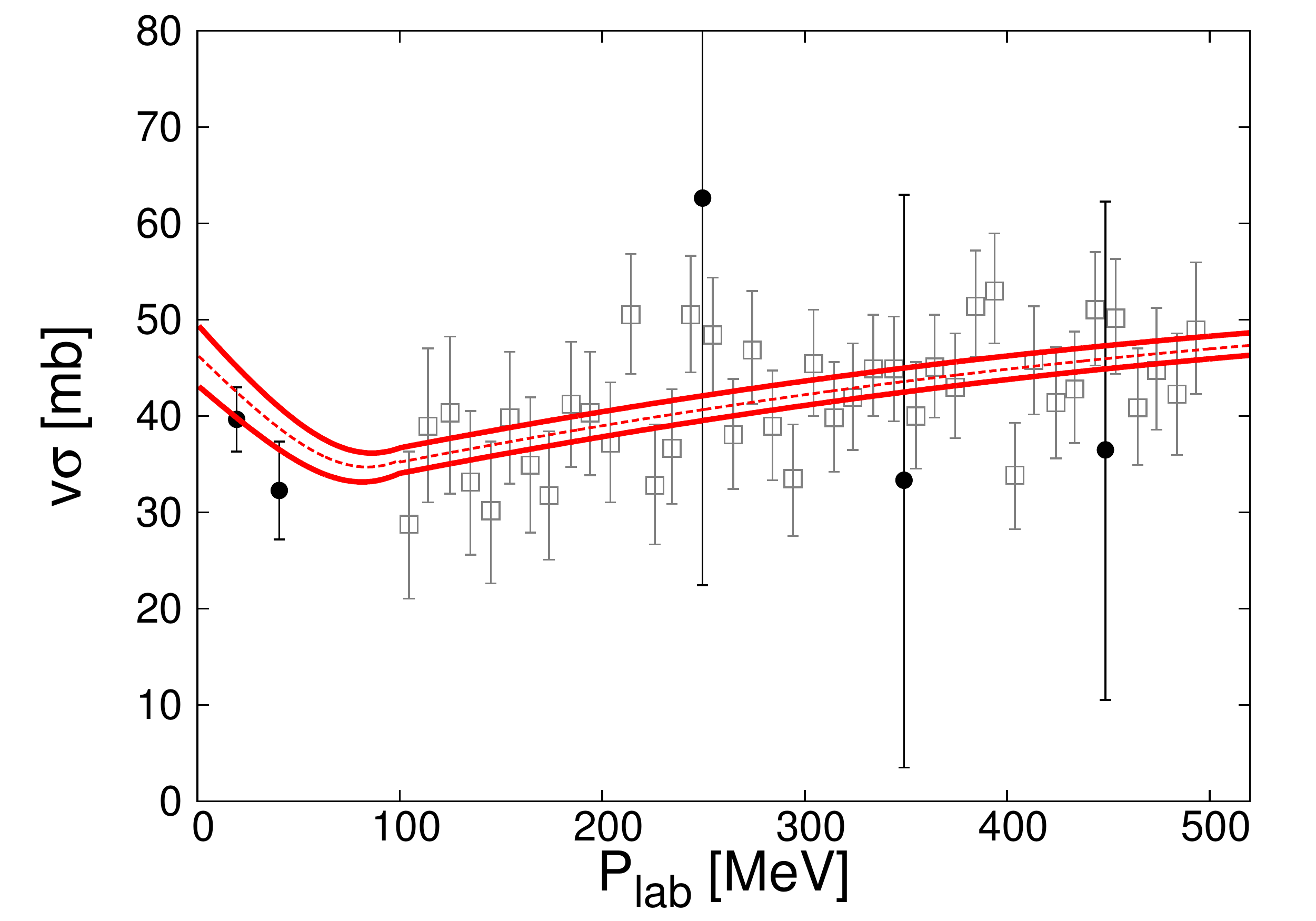}
\caption{\small \label{fig:ann} 
A fit to the data in Fig.~14 of Ref.~\cite{Mutchler:1988av}. 
The red dashed curve is the best fit with $\chi^2/{\rm dof}=0.74$,
and the red solid curves show the 1$\sigma$ allowed range.
The annihilation cross section at zero momentum from our fit is consistent with that of Ref.~\cite{Mutchler:1988av}:
$v\sigma=44\pm 3.5$~mb.
}
\end{figure}

According to Ref.~\cite{Mutchler:1988av},
 $v\sigma(\bar{n}n \to {\rm pions})=v\sigma(\bar{n}p\to {\rm pions})_{\rm exp}=44$~mb at zero momentum, where we have assumed that the $\bar{n}n$ annihilation cross section is the same as the $\bar{n}p$ cross section.
We checked this result by fitting the data in Ref.~\cite{Mutchler:1988av}; see Fig.~\ref{fig:ann}.
 Using the branching ratios, 
${\rm Br}(\bar{n}n\to \pi^+\pi^-\pi^0)=0.065$
and ${\rm Br}(\bar{n}n\to (\pi^+\pi^-3\pi^0)+(2\pi^+2\pi^-\pi^0))=0.28+0.24=0.52$~\cite{Abe:2011ky}, which were obtained from $\bar{p}p$ and $\bar{p}d$ bubble chamber data, we find the numerical values of
$\sigma(\bar{n}n \to 3\pi^0)=0.065 \cdot \sigma(\bar{n}n \to {\rm pions})$ 
and 
$\sigma(\bar{n}n \to 5\pi^0)=0.52 \cdot \sigma(\bar{n}n \to {\rm pions})$. 
Next, we determine $y$ by comparing our Monte Carlo calculations with
the values for $\sigma(\bar{n}n \to 3\pi^0)$ 
and $\sigma(\bar{n}n \to 5\pi^0)$.
To calculate the leading order cross sections, 
we implement the neutron-pion interaction in the CalcHEP package~\cite{calchep}, generate the amplitude squared, 
and insert the form factor of Eq.~(\ref{eq:ff}) by hand.
To evaluate $\sigma(\bar{n}n\to 3\pi^0)(y)$ 
and $\sigma(\bar{n}n\to 5\pi^0)(y)$, where we have made the $y$-dependence explicit, 
we perform the three-body and five-body phase space integrations, respectively.  
Matching the cross sections,
$\sigma(\bar{n}n\to 3\pi^0)(y)=\sigma(\bar{n}n \to 3\pi^0)$
and $\sigma(\bar{n}n\to 5\pi^0)(y)=\sigma(\bar{n}n \to 5\pi^0)$, yields $y=0.542$ and $y=0.337$, respectively. 

Using the above values of $y$, we  calculate
$\sigma(\bar{\chi}n\to \phi 3\pi^0)$
and $\sigma(\bar{\chi}n\to \phi 5\pi^0)$ for Model II,
by replacing the initial particle $\bar{n}$ by $\bar{\chi}$,
multiplying by the mixing parameter $\theta$, 
and including the dark sector particle $\phi$ in the final state.
We implement the $\chi$-$n$-$\phi$ interaction in CalcHEP,
output the amplitude squared, convolve with the form factor, and perform the four-body and six-body phase space integrations.

\end{document}